\documentclass[letterpaper,aps,prd,twocolumn,
tightenlines,preprintnumbers,
nofootinbib,showkeys,
superscriptaddress,longbibliography]{revtex4-2}

\usepackage{multirow}
\usepackage[T1]{fontenc}
\usepackage{XCharter}          
\usepackage{bm,bbold}          
\usepackage{amsmath,amssymb}   
\usepackage{siunitx}           
\usepackage{slashed}          

\usepackage{graphicx}         
\usepackage{subfig}          
 
\usepackage{booktabs,multirow,makecell}  %
\usepackage{float}            
\usepackage[export]{adjustbox} 
\usepackage[left=1.58cm,right=1.58cm,
top=1.83cm,bottom=1.84cm]{geometry}
\usepackage[dvipsnames]{xcolor}
\usepackage{hyperref}
\hypersetup{
	colorlinks=true,
	citecolor=red,
	linkcolor=NavyBlue,
	urlcolor=blue
}
\usepackage{natbib}
\usepackage{relsize}

\newcommand{\orcid}[1]{\hspace{1mm}\href{https://orcid.org/#1}{\includegraphics[height=0.3cm,keepaspectratio]{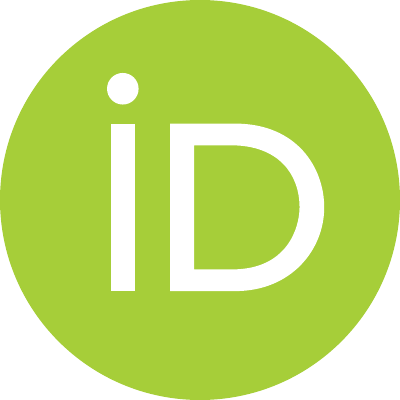}}}

\begin{document}
\preprint{IFJPAN-IV-2026-4}

\title{Probing doubly charged Higgs bosons \\[0.15cm] with three-body associated production at future $e^+e^-$ colliders}
\relscale{0.9}
\author{B. Ait-Ouazghour\orcid{0009-0006-1419-969X}}
\email{b.ouazghour@gmail.com}
\affiliation{Cadi Ayyad University,  Faculty of Science Semlalia,  LPHEAG, P.O.B. 2390 Marrakech 40000, Morocco.}
\author{A. Arhrib\orcid{0000-0001-5619-7189}}
\email{aarhrib@gmail.com}
\affiliation{Abdelmalek Essaadi University, FST Tanger B.P. 416, Morocco}
\affiliation{LAPTh, CNRS, Universit\'e Savoie Mont-Blanc, 9 Chemin de Bellevue, 74940, Annecy, France.}

\author{R. Benbrik\orcid{0000-0002-5159-0325}}
\email{r.benbrik@uca.ac.ma}
\affiliation{Laboratory of Physics, Energy, Environment, and Applications, Cadi Ayyad University, Sidi Bouzid, P.O. Box 4162, Safi, Morocco.}

\author{M. Boukidi\orcid{0000-0001-9961-8772}}
\email{mohammed.boukidi@ifj.edu.pl}
\affiliation{Institute of Nuclear Physics, Polish Academy of Sciences, ul. Radzikowskiego 152, Cracow, 31-342, Poland.}

\author{M. Chabab\orcid{0000-0002-2772-4290}}
\email{mchabab@uca.ac.ma}
\affiliation{Cadi Ayyad University,  Faculty of Science Semlalia,  LPHEAG, P.O.B. 2390 Marrakech 40000, Morocco.}
\affiliation{Cadi Ayyad University, National School of Applied Science, P.O.B. 63 Safi 46000, Morocco}
\author{K. Goure\orcid{0009-0007-5292-5012}}
\email{khalidgoure01@gmail.com}
\affiliation{Cadi Ayyad University,  Faculty of Science Semlalia, LPHEAG, P.O.B. 2390 Marrakech 40000, Morocco.}

\author{S. Moretti\orcid{0000-0002-8601-7246}}
\email{stefano.moretti@physics.uu.se/s.moretti@soton.ac.uk}
\affiliation{Department of Physics and Astronomy, Uppsala University, Box 516, SE-751 20 Uppsala, Sweden.}
\affiliation{School of Physics and Astronomy, University of Southampton, Southampton, SO17 1BJ, United Kingdom.}
\date{\today}

\begin{abstract}
	We study the discovery prospects for a doubly charged Higgs boson $H^{\pm\pm}$ in the 2-Higgs doublet model with type-II seesaw at future $e^+e^-$ colliders. Focusing on the three-body channels $e^+e^- \to H^{\pm\pm}H_1^{\mp}H_1^{\mp}$ and $e^+e^- \to H^{\pm\pm}H_1^{\mp}W^{\mp}$, we scan the model parameter space subject to theoretical consistency as well as current collider, flavor and electroweak precision observables (EWPOs). We find that these $2\to3$ production modes can exceed the conventional pair-production rate $e^+e^- \to H^{++}H^{--}$, followed by $H^{\pm\pm}\to H_1^{\pm}H_1^{\pm}$ and $H^{\pm}_1W^{\pm}$ decays, over wide regions, particularly above the $H^{\pm\pm}\to H_1^{\pm}H_1^{\pm}$ and $H^{\pm\pm}\to H_1^{\pm}W^{\pm}$ thresholds, reaching cross sections up to ${\cal O}(10^2)$~fb for $\sqrt{s}=500$--$1500$~GeV. A detector-level analysis of the $4\ell+\slashed{E}_T$ signature, including dominant multiboson and top quark backgrounds, shows that discovery sensitivity is achievable for $\sqrt{s}=1000$-$1500$~GeV with integrated luminosities in the few ab$^{-1}$ range, even in the presence of realistic systematic uncertainties.
\end{abstract}
\maketitle

\section{Introduction}

Despite its remarkable success, the Standard Model (SM) leaves a number of fundamental questions unanswered, including the nature of Dark Matter (DM), the origin of the baryon asymmetry of the Universe, and the stability of the EW  scale~\cite{Zwicky:1933gu,Rubin:1970zza,Veltman:1980mj,SupernovaSearchTeam:1998fmf,Crivellin:2023zui}. These open issues have motivated extensive efforts to go Beyond the SM (BSM), with a particularly well-motivated direction being an extended Higgs sector. In parallel, the observation of neutrino oscillations established that neutrinos are massive and mix, thereby requiring new BSM dynamics~\cite{Kajita:2016cak,McDonald:2016ixn,Cai:2017mow}. In this respect, a broad class of possible explanations is provided by seesaw mechanisms, which generate tiny neutrino masses from the dimension-five Weinberg operator~\cite{Weinberg:1979sa}. Besides accounting for neutrino masses and the lepton-mixing pattern, seesaw frameworks often predict lepton-number-violating signatures and new states that can be searched for at collider experiments.

The Weinberg operator can be Ultra-Violet (UV) completed through three canonical realizations~\cite{Weinberg:1979sa}: the type-I seesaw, featuring an $SU(2)_L$-singlet fermion~\cite{Minkowski:1977sc,Gell-Mann:1979vob,Mohapatra:1979ia,Schechter:1980gr}, the type-II seesaw, based on an $SU(2)_L$-triplet scalar~\cite{Cheng:1980qt,Schechter:1980gr,Mohapatra:1980yp}, and the type-III seesaw, involving an $SU(2)_L$-triplet fermion~\cite{Foot:1988aq}. In the type-II case, light neutrino masses are induced once the neutral component of the triplet develops a Vacuum Expectation Value (VEV), such that $m_\nu \propto v_t\,Y_\Delta$, where $v_t$ denotes the triplet VEV and $Y_\Delta$ the corresponding Yukawa couplings.

A characteristic prediction of the type-II seesaw is the presence of a doubly charged Higgs boson, $H^{\pm\pm}$. The phenomenology of $H^{\pm\pm}$ states has been studied extensively at hadron and lepton colliders based  on a variety of theoretical contexts~\cite{Melfo:2011nx,Huitu:1996su,Gunion:1996pq,Chakrabarti:1998qy,Muhlleitner:2003me,Chun:2003ej,Akeroyd:2005gt,Han:2007bk,Akeroyd:2007zv,delAguila:2008cj,Perez:2008ha,Akeroyd:2009hb,Akeroyd:2010ip,Aoki:2011pz,Arhrib:2011uy,Akeroyd:2011zza,Chiang:2012dk,Sugiyama:2012yw,Akeroyd:2012nd,Chun:2012zu,delAguila:2013mia,Chun:2013vma,Kanemura:2013vxa,Kanemura:2014goa,Kanemura:2014ipa,Dutta:2014dba,Kang:2014lwn,Han:2015hba,Han:2015sca,Mitra:2016wpr,Ghosh:2017pxl,Antusch:2018svb,BhupalDev:2018tox,deMelo:2019asm,Primulando:2019evb,Chun:2019hce,Fuks:2019clu,Padhan:2019jlc,Ashanujjaman:2021txz,Ashanujjaman:2022ofg,Cakir:2006pa,Nomura:2017abh,Blunier:2016peh,Crivellin:2018ahj,Agrawal:2018pci,Ashanujjaman:2022tdn,Ruiz:2022sct,Chiang:2025lab,Dev:2019hev,Yang:2021skb,Deppisch:2015qwa,Li:2023ksw,Maharathy:2023dtp,Jueid:2023qcf,Belfkir:2023lot,Jia:2024wqi,	CMS:2012dun,ATLAS:2012hi,CMS:2014mra,BhupalDev:2013xol,ATLAS:2014kca,ATLAS:2014vih,ATLAS:2017xqs,CMS:2017fhs,ATLAS:2021jol,ATLAS:2023sua,ATLAS:2024txt,Guedes:2025jqu}.
Importantly, the collider signatures of $H^{\pm\pm}$ are largely controlled by $v_t$, which is simultaneously the key parameter governing neutrino-mass generation. For $v_t \ll 10^{-4}\,\text{GeV}$, the doubly charged Higgs boson decays predominantly into same-sign dileptons, $H^{\pm\pm}\to \ell^\pm\ell^\pm$ ($\ell=e,\mu$)~\cite{Han:2007bk,Perez:2008ha}. Conversely, for $v_t \gg 10^{-4}\,\text{GeV}$, the decay into same-sign dibosons, $H^{\pm\pm}\to W^\pm W^\pm$, becomes dominant~\cite{Han:2007bk,Perez:2008ha}. Consequently, present and future searches typically target both dilepton and diboson final states.

In this work we focus on the 2-Higgs doublet model with type-II seesaw~(2HDMcT)~\cite{Chen:2014xva,Ouazghour:2018mld,Ouazghour:2023eqr,Ouazghour:2024fgo,AitOuazghour:2025cmd,BrahimAit-Ouazghour:2024img,BrahimAit-Ouazghour:2025mhy}. Since mass generation in seesaw frameworks is intimately tied to EW Symmetry Breaking (EWSB), in close analogy to the Brout--Englert--Higgs mechanism, scalar-extended realizations are especially appealing and lead to a rich phenomenology. Relative to the standard 2HDM scalar sector, the 2HDMcT exhibits a broader spectrum and distinctive signatures, with $H^{\pm\pm}$ providing a particularly clean ``smoking gun'.'  Furthermore, the 2HDMcT can be viewed as one of the simplest and most economical extensions of the SM that can simultaneously accommodate neutrino masses~\cite{Perez:2008ha,King:2015aea} and address the DM problem~\cite{Chen:2014lla}. 

A doubly charged Higgs  state is currently being intensively searched for by the ATLAS and CMS Collaborations at the Large Hadron Collider (LHC), notably, through pair production $pp\to Z/\gamma\to H^{++}H^{--}$ followed by same-sign dilepton decays~\cite{Ouazghour:2018mld}. A key quantitative difference between the 2HDMcT and the minimal Higgs-Triplet Model (HTM) emerges in certain cascade topologies.  In the 2HDMcT the chain $pp \to Z/\gamma\to H_2^{+}H_2^{-}\to H^{++}W^{-}\,H^{--}W^{+}\to\ell^{+}\ell^{+}\ell^{-}\ell^{-}+4j$ can proceed fully on shell. This is because the additional scalars relax the oblique-parameter bounds and allow large mass splittings $\Delta m = m_{H^\pm} - m_{H^{\pm\pm}}$ reaching values of order ${\cal O}(m_W)$ for $H_2^\pm$. In the HTM, electroweak precision data confine $|\Delta m|\lesssim40$ GeV\,\cite{Melfo:2011nx,Arhrib:2011uy,Ashanujjaman:2021txz}, forcing off-shell $W$ bosons\footnote{The sign of $\Delta m$ can be positive or negative in the HTM\,(See Ref.~\cite{Ashanujjaman:2025scr}).} in the analogous sequence $pp\to Z/\gamma\to H^{+}H^{-}\to H^{++}W^{-*}\,H^{--}W^{+*}$ and severely suppressing its rate. The larger mass splitting in the 2HDMcT opens the on-shell decay channels $H^{\pm\pm}\to W^{\pm}H^{\pm}_{1}$ and $H^{\pm\pm}\to H^{\pm}H^{\pm}_{1}$, thereby rendering the associated production modes $H^{\pm\pm}H^{\mp}_{1}H^{\mp}_{1}$ and $H^{\pm\pm}H^{\mp}_{1}W^{\mp}$ promising discovery channels for $H^{\pm\pm}$ at current and future colliders. Beyond collider phenomenology, it has also been shown that interactions between the doublet and triplet fields can induce a Strong First-Order EW Phase Transition (SFOEWPT), thereby enabling EW  baryogenesis~\cite{Ramsey-Musolf:2019lsf}. Overall, compared to the HTM, the production and decay patterns of $H^{\pm\pm}$ in the 2HDMcT are significantly modified due to the presence of the second Higgs doublet (see Ref.~\cite{Chen:2014qda}).

In this study, we investigate doubly charged Higgs production in association with a pair of singly charged Higgs bosons,
$e^+ e^- \to H^{\pm\pm} H_1^\mp H_1^{\mp}$,
as well as its production  in association with a $W^\pm$ boson and a singly charged Higgs state,
$e^+ e^- \to H^{\pm\pm} H_1^\mp W^{\mp}$,
within the type-X 2HDMcT framework at future electron-positron colliders. Our numerical results are obtained from a detailed scan of its parameter space, subject to theoretical requirements (perturbative unitarity and vacuum stability) and current experimental constraints (including measurements of the SM-like Higgs boson, exclusion limits on additional Higgs states, EWPOs, flavor bounds and limits from lepton-flavor-violating processes). We perform a full Monte Carlo (MC) simulation at detector level and quantify the sensitivity to these channels at the, e.g., International Linear Collider (ILC) \cite{ILCInternationalDevelopmentTeam:2022izu} for Center-of-Mass (CoM) energies of $500, 1000$ and $1500~\text{GeV}$\footnote{The choice of the ILC is merely related to the use of a detector card specific to such a machine, while our results are equally applicable to other TeV scale collider setups, like the Compact Linear Collider (CLIC)~\cite{Linssen:2012hp}.}.

This paper is organized as follows. In Sect.~\ref{prese_2HDMcT}, we briefly review the 2HDMcT model. Sect.~\ref{constraint} summarizes the theoretical and experimental constraints imposed on the parameter space. Our computational procedure and numerical results are presented in Sect.~\ref{COMPUTATIONAL_PROCEDURE}. Finally, Sect.~\ref{conlusion} is devoted to our conclusions.

\section{2HDMcT: BRIEF REVIEW}

\label{prese_2HDMcT}

The 2HDMcT contains two Higgs doublets $\Phi_{i}$ ($i = 1,2$)	and one colorless Higgs field $\Delta$ transforming as a triplet under the $SU(2)_L$ gauge group with hypercharge $Y_\Delta=2$. In this case, the most general gauge-invariant Lagrangian of the  2HDMcT is given by 

	\begin{equation}
		\begin{matrix}
			\mathcal{L}=\sum_{i=1}^2(D_\mu{\Phi_i})^\dagger(D^\mu{\Phi_i})+Tr(D_\mu{\Delta})^\dagger(D^\mu{\Delta})\vspace*{0.12cm}\\
			\hspace{-3cm}-V(\Phi_i, \Delta)+\mathcal{L}_{\rm Yukawa},
			\label{eq:thdmt-lag}
		\end{matrix}
	\end{equation}
	where the covariant derivatives are defined as
	\begin{equation}
		D_\mu{\Phi_i}=\partial_\mu{\Phi_i}+igT^a{W}^a_\mu{\Phi_i}+i\frac{g'}{2}B_\mu{\Phi_i}, \label{eq:covd1}
	\end{equation}
	\vspace*{-0.6cm}
	\begin{equation}
		D_\mu{\Delta}=\partial_\mu{\Delta}+ig[T^a{W}^a_\mu,\Delta]+ig' \frac{Y_\Delta}{2} B_\mu{\Delta}, \label{eq:covd2}
	\end{equation}
with 
	(${W}^a_\mu$, $g$) and ($B_\mu$, $g'$) denoting, respectively, the $SU(2)_L$ and $U(1)_Y$ (gauge fields, couplings) while  $T^a \equiv \sigma^a/2$ where the $\sigma^a$ ($a=1, 2, 3$)  are the Pauli matrices. In terms of the two $SU(2)_L$ Higgs doublets $\Phi_i$ and the triplet field $\Delta$, the 2HDMcT Higgs potential is given by \cite{Chen:2014xva,Ouazghour:2018mld}:
\begin{widetext}
	\begin{equation}
		\begin{aligned}
			&V(\Phi_1,\Phi_2,\Delta) = m_{11}^2 \Phi_1^\dagger\Phi_1
			+ m_{22}^2\Phi_2^\dagger\Phi_2
			- \left[m_{12}^2\Phi_1^\dagger\Phi_2 + {h.c.}\right]
			+ \frac{\lambda_1}{2}(\Phi_1^\dagger\Phi_1)^2
			+ \frac{\lambda_2}{2}(\Phi_2^\dagger\Phi_2)^2
			+\lambda_3(\Phi_1^\dagger\Phi_1)(\Phi_2^\dagger\Phi_2)+\lambda_4(\Phi_1^\dagger\Phi_2)(\Phi_2^\dagger\Phi_1)
			\\
			&\quad
			+ \left\{
			\frac{\lambda_5}{2}(\Phi_1^\dagger\Phi_2)^2
			+ \big[\beta_1(\Phi_1^\dagger\Phi_1)
			+\beta_2(\Phi_2^\dagger\Phi_2)\big]\Phi_1^\dagger\Phi_2+ {h.c.}\right\} + \lambda_6\,\Phi_1^\dagger \Phi_1 \, \mathrm{Tr}(\Delta^{\dagger}\Delta)
			+ \lambda_7\,\Phi_2^\dagger \Phi_2 \, \mathrm{Tr}(\Delta^{\dagger}\Delta)
			+ \lambda_8\,\Phi_1^\dagger \Delta \Delta^{\dagger} \Phi_1
			\\
			&\quad
			+ \lambda_9\,\Phi_2^\dagger \Delta \Delta^{\dagger} \Phi_2
			+ m^2_{\Delta}\, \mathrm{Tr}(\Delta^{\dagger}\Delta)
			+ \bar{\lambda}_8 \left(\mathrm{Tr}\,\Delta^{\dagger}\Delta\right)^2
			+ \bar{\lambda}_9 \, \mathrm{Tr}\left(\Delta^{\dagger}\Delta\right)^2
			+ \left\{\mu_1 \Phi_1^T i\sigma^2 \Delta^{\dagger}\Phi_1 + \mu_2 \Phi_2^T i\sigma^2 \Delta^{\dagger}\Phi_2 + \mu_3 \Phi_1^T i\sigma^2 \Delta^{\dagger}\Phi_2 + {h.c.}\right\}
		\end{aligned}
		\label{scalar_pot}
	\end{equation}
\end{widetext}

	where $Tr$ denotes the trace over 2x2 matrices. The triplet $\Delta$ and the Higgs doublets $\Phi_{i}$ are represented by 
	\begin{eqnarray}
		\Delta &=&\left(
		\begin{array}{cc}
			\delta^+/\sqrt{2} & \delta^{++} \\
			(v_t+\delta^0+i\eta_0)/\sqrt{2} & -\delta^+/\sqrt{2}\\
		\end{array}
		\right)\end{eqnarray}
	\vspace*{-0.25cm}
	\begin{eqnarray}
		\Phi_1&=&\left(
		\begin{array}{c}
			\phi_1^+ \\
			\phi^0_1 \\
		\end{array}
		\right){,}~~~\Phi_2=\left(
		\begin{array}{c}
			\phi_2^+ \\
			\phi^0_2 \\
		\end{array}
		\right)\end{eqnarray}
with $\phi^0_1=(v_1+\rho_1+i\eta_1)/\sqrt{2}$ and $\phi^0_2=(v_2+\rho_2+i\eta_2)/\sqrt{2}$. After spontaneous EWSB, the Higgs doublets and triplet fields acquire their VEVs denoted by, respectively, $v_1$, $v_2$ and $v_t$, so that eleven physical Higgs states appear, namely,  three CP-even neutral Higgs bosons $(h_1, h_2, h_3)$, four singly charged Higgs bosons\footnote{A detailed study of the singly charged Higgs boson at muon colliders was performed in Ref.~\cite{BrahimAit-Ouazghour:2025mhy}.} $(H_1^{\pm}, H_2^{\pm})$,  two CP-odd neutral Higgs bosons $(A_1, A_2)$, and, finally, two doubly charged Higgs bosons $H^{\pm\pm}$.  

Expanding the covariant derivative ${\rm D}_\mu$, and performing the usual transformations on the gauge and scalar fields to obtain the physical fields, one can identify the Higgs couplings of $h_i$ to the massive gauge bosons $V=W,Z$ as given in Tab.~\ref{table2}. Note that in our model, the triplet field $\Delta$ couples directly to the SM particles, so a new contribution will appear, and the two couplings $C^{h_i}_V$ ($V=W^\pm, Z$) differs from one to another by a factor 2 associated to $v_t$.
\subsection{The Yukawa sector}
The Yukawa Lagrangian encompasses the entire Yukawa sector of the 2HDM along with an additional term $\mathcal{L}_{\rm Yukawa}$. We list in Tab.~\ref{table1} all the CP-even $h_i$ ($i=1,2,3$) and CP-odd $A_j$ ($j=1,2$) Yukawa couplings for the type-X Yukawa texture in the model.
The additional Yukawa term responsible for neutrino masses is given by
\begin{eqnarray}
	-\mathcal{L}_{\rm Yukawa} = Y_{ij}L_i^TCi\sigma_2\Delta L_j + h.c.,
	\label{neutrino_mass}
\end{eqnarray}
where $Y$ is a $3\times3$ complex symmetric matrix and $L = (\nu_L,l_L)^T$ denotes the $SU(2)_L$ doublets of left-handed leptons. This term, following EWSB, generates tiny (Majorana) masses for the neutrinos, which can be obtained from the Yukawa terms in Eq.\,\ref{neutrino_mass} as
\begin{eqnarray}
	m_\nu = \sqrt{2}\,Y v_\Delta.
\end{eqnarray}
The $3 \times 3$ neutrino mass matrix, $m_\nu$, can then be diagonalized through a unitary 
transformation involving the Pontecorvo--Maki--Nakagawa--Sakata (PMNS) matrix, 
$U_{\mathrm{PMNS}}$, according to
\begin{equation}
	U_{\mathrm{PMNS}}^T \, m_\nu \, U_{\mathrm{PMNS}} 
	= m_\nu^d 
	= \mathrm{diag}(m_1, m_2, m_3) \, ,
\end{equation}
where $m_1, m_2$, and $m_3$ are the three neutrino mass eigenvalues. 
The matrix $U_{\mathrm{PMNS}}$ is conventionally parameterized in terms of 
three mixing angles ($\theta_{12}, \theta_{23}, \theta_{13}$), 
one Dirac CP-violating phase ($\delta$), 
and two Majorana phases ($\phi_1, \phi_2$): it is given by
\begin{widetext}

\begin{equation}
	U_{\rm PMNS} =
	\begin{pmatrix}
		c_{12} c_{13} & s_{12} c_{13} & s_{13} e^{-i\delta} \\[6pt]
		- c_{12} s_{13} s_{23} e^{i\delta} - c_{23} s_{12} & 
		c_{12} c_{23} - s_{12} s_{13} s_{23} e^{i\delta} & 
		c_{13} s_{23} \\[6pt]
		s_{12} s_{23} - c_{12} c_{23} s_{13} e^{i\delta} & 
		- c_{23} s_{12} s_{13} e^{i\delta} - c_{12} s_{23} & 
		c_{13} c_{23}
	\end{pmatrix}
	\cdot
	\mathrm{diag}\!\left(e^{i\Phi_1/2},\, 1,\, e^{i\Phi_2/2}\right).
\end{equation}
\end{widetext}

Assuming a Normal Hierarchy (NH) of neutrino masses, $m_{\nu_1} < m_{\nu_2} < m_{\nu_3}$, all free parameters of the neutrino sector can be expressed in terms of the neutrino oscillation parameters and the mass of the lightest neutrino:
\begin{equation}
	 m_{\nu_1},  \ \theta_{12}, \ \theta_{13}, \ \theta_{23},\ \Delta m_{21}^2, \ \Delta m_{31}^2, \ \delta, \ \phi_1, \ \phi_2.
\end{equation}
The neutrino mass-squared differences in the NH are taken as $\Delta m_{31}^2>0$ and $\Delta m_{21}^2>0$, allowing one to determine the heavier neutrino masses through
\begin{equation}
	m_{\nu_2} = \sqrt{m_{\nu_1}^2 + \Delta m_{21}^2}, \qquad 
	m_{\nu_3} = \sqrt{m_{\nu_1}^2 + \Delta m_{31}^2}.
\end{equation}

In the case of an Inverted Hierarchy (IH), the lightest neutrino is $m_{\nu_3}$ ($m_{\nu_3} < m_{\nu_2} < m_{\nu_1}$), which sets the overall neutrino mass scale. The corresponding set of input parameters becomes
\begin{equation}
	 m_{\nu_3}, \ \theta_{12}, \ \theta_{13}, \ \theta_{23}, \ \Delta m_{21}^2, \ \Delta m_{32}^2, \ \delta, \ \phi_1, \ \phi_2  \,,
\end{equation}
with $\Delta m_{21}^2>0$ and $\Delta m_{32}^2<0$, with  the remaining neutrino masses determined by
\begin{equation}
	m_{\nu_1} = \sqrt{m_{\nu_3}^2 - \Delta m_{32}^2 - \Delta m_{21}^2},
	m_{\nu_2} = \sqrt{m_{\nu_3}^2 - \Delta m_{32}^2}.
\end{equation}
The parameters $m_{\nu_3}, \ \Delta m_{21}^2, \ \Delta m_{32}^2, \ \theta_{12}, \ \theta_{13}, \ \theta_{23}, \delta$ are largely fixed by the global fit of oscillation data~\cite{Esteban:2024eli}, 
while the smallest neutrino mass $m_{\nu_{\min}}$ = $m_{\nu_1}(m_{\nu_3})$ in NH (IH), and the two Majorana phases $\phi_1, \ \phi_2$ are not constrained  by current oscillation data. 
	\begin{table*}[t]
		\begin{center}
			\begin{tabular}{|c|c|c|}
				\hline  
				& $C^{h_i}_W$    & $C^{h_i}_Z$  \\
				\hline  $h_1$ & $\displaystyle{\frac{v_1}{v} \mathcal{E}_{11} + \frac{v_2}{v} \mathcal{E}_{21} + 2\,\frac{v_t}{v} \mathcal{E}_{31}}$ & 
				$\displaystyle{\frac{v_1}{v} \mathcal{E}_{11} + \frac{v_2}{v} \mathcal{E}_{21} + 4\,\frac{v_t}{v} \mathcal{E}_{31}}$  \\
				\hline  $h_2$ & $\displaystyle{\frac{v_1}{v} \mathcal{E}_{12} + \frac{v_2}{v} \mathcal{E}_{22} + 2\,\frac{v_t}{v} \mathcal{E}_{32}}$ & 
				$\displaystyle{\frac{v_1}{v} \mathcal{E}_{12} + \frac{v_2}{v} \mathcal{E}_{22} + 4\,\frac{v_t}{v} \mathcal{E}_{32}}$  \\
				\hline  $h_3$ & $\displaystyle{\frac{v_1}{v} \mathcal{E}_{13} + \frac{v_2}{v} \mathcal{E}_{23} + 2\,\frac{v_t}{v} \mathcal{E}_{33}}$ & 
				$\displaystyle{\frac{v_1}{v} \mathcal{E}_{13} + \frac{v_2}{v} \mathcal{E}_{23} + 4\,\frac{v_t}{v} \mathcal{E}_{33}}$  \\
				\hline 
			\end{tabular}
			\caption{The normalized couplings of the neutral CP-even Higgs bosons ($h_i$)
				to the  gauge bosons $V=W,Z$ in the 2HDMcT.}
			\label{table2}
		\end{center}
	\end{table*}
	
In the charged sector, the doubly charged eigenvalue $m_{H^{\pm\pm}}^2$, corresponding to the doubly charged eigenstate $H^{\pm\pm}$, can simply be determined by collecting all the coefficients of $\delta^{++}\delta^{--}$ in the Higgs potential. This gives, 
	\begin{widetext}
	\begin{eqnarray}
		m_{H^{\pm\pm}}^2=\frac{\sqrt{2}\mu_1 v_1^2 + \sqrt{2}\mu_3 v_1 v_2 + \sqrt{2}\mu_2 v_2^2 - \lambda_8 v_1^2 v_t
			- \lambda_9 v_2^2 v_t - 2 \bar{\lambda}_9 v_t^3}{2v_t}.  \label{eq:mHpmpm}
	\end{eqnarray}
    \end{widetext}
	   \begin{table*}[t]
	\begin{center}
		\begin{tabular}{|c|c|c|c|c|c|c|c|c|c|c|c|c|c|c|c|c|c}
			\hline  & $C^{h_1}_U$    & $C^{h_1}_D$ & $C^{h_1}_\ell$   &   $C^{h_2}_U$   &   $C^{h_2}_D$ &   $C^{h_2}_\ell$   &   $C^{h_3}_U$  &   $C^{h_3}_D$ &   $C^{h_3}_\ell$  & $C^{A_1}_U$  & $C^{A_1}_D$ & $C^{A_1}_\ell$ & $C^{A_2}_U$  & $C^{A_2}_D$ & $C^{A_2}_\ell$ \\
			\hline  Type-X &$\displaystyle\frac{\mathcal{E}_{12}}{s_\beta} $& $\displaystyle\frac{\mathcal{E}_{12}}{s_\beta} $ &
			$\displaystyle\frac{\mathcal{E}_{11}}{c_\beta} $ & $\displaystyle\frac{\mathcal{E}_{22}}{s_\beta} $ & $\displaystyle\frac{\mathcal{E}_{22}}{s_\beta} $ &
			$\displaystyle\frac{\mathcal{E}_{21}}{c_\beta} $ & $\displaystyle\frac{\mathcal{E}_{32}}{s_\beta} $ &$\displaystyle\frac{\mathcal{E}_{32}}{s_\beta} $
			&$\displaystyle\frac{\mathcal{E}_{31}}{c_\beta} $&
			$\displaystyle\frac{\mathcal{O}_{22}}{s_\beta} $&
			$\displaystyle\frac{\mathcal{O}_{22}}{s_\beta} $&
			$\displaystyle\frac{\mathcal{O}_{21}}{c_\beta} $ &
			$\displaystyle\frac{\mathcal{O}_{32}}{s_\beta} $& $\displaystyle\frac{\mathcal{O}_{32}}{s_\beta} $& 
			$\displaystyle\frac{\mathcal{O}_{31}}{c_\beta} $ \\

			\hline
		\end{tabular}
	\end{center} 
	\caption{The normalized Yukawa couplings coefficients of the neutral Higgs bosons ($h_i$ and $A_j$) to the leptons ($\ell$), up ($U$),  and down quarks ($D$) in 2HDMcT. For the expressions of the coefficients $\displaystyle{\mathcal{E}_{ij}}$ and  $\displaystyle{\mathcal{O}_{ij}}$, see Ref~\cite{Ouazghour:2018mld}.}
	\label{table1}
\end{table*}
\vspace{6pt}
	\section{Constraints}
\vspace{6pt}
\label{constraint}
\paragraph*{}
The phenomenological analysis in 2HDMcT is performed via  implementation of a full set of theoretical constraints ~\cite{Ouazghour:2018mld, Ouazghour:2023eqr} as well as Higgs exclusion limits from various experimental measurements at colliders, as follows.
\begin{itemize}
	\item \textbf{Unitarity}: The Higgs and Goldstone $2\to2$ scattering matrix elements must obey conservation of probability.
	\item \textbf{Perturbativity}: The quartic couplings of the Higgs potential are constrained by the following conditions: $| \lambda_i|<8 \pi$.
	\item \textbf{Vacuum Stability}: Boundedness from Below (BFB)  arising from the positivity in any direction of the fields $\Phi_i$ and $\Delta$ is required.
	\item[\textbullet]{\bf EWPOs}: The oblique parameters $S, T$, and $U$~\cite{Peskin:1991sw,Grimus:2008nb} have been calculated in the 2HDMcT~\cite{Ouazghour:2023eqr} and measured at the Large Electron-Positron (LEP) collider as well as Tevatron. In the  light of the new Particle Data Group (PDG) mass of the $W^\pm$ boson, they yield~\cite{ParticleDataGroup:2024cfk} (including correlation):	
	\begin{eqnarray}
		\widehat S_0= -0.05\pm 0.07,  \widehat T_0 = 0.00\pm 0.06, \rho_{ST} = 0.93,  \nonumber 
	\end{eqnarray}	
	for which we use the following $\chi^2_{ST}$ test:
	\begin{equation}
		\small
		\label{eq:STRange}
		\frac{(S-\widehat S_0)^2}{\sigma_S^2}\ +\
		\frac{(T-\widehat T_0)^2}{\sigma_T^2}\ -\
		2\rho_{ST}\frac{(S-\widehat S_0)(T-\widehat T_0)}{\sigma_S \sigma_T}\
		\leq\ R^2\,(1-\rho_{ST}^2)\; ,
	\end{equation}
	with $R^2=2.3$ and $5.99$ corresponding to $68.3 \%$  and
	$95 \% $  Confidence Level (CL), respectively.
	Our numerical analysis is performed with $\chi^2_{ST}$ at 95\% CL. 
	\item \textbf{Colliders}: To further delimit the allowed parameter space, the {\texttt{HiggsTools}} package~\cite{Bahl:2022igd} is employed. This ensures that the allowed parameter regions align with the observed properties of the $125$~GeV Higgs boson  (via \texttt{HiggsSignals}~\cite{Bechtle:2013xfa,Bechtle:2014ewa,Bechtle:2020uwn,Bahl:2022igd}) and with the limits from searches for additional Higgs bosons at the LHC, Tevatron and LEP (via \texttt{HiggsBounds}~\cite{Bechtle:2008jh,Bechtle:2011sb,Bechtle:2013wla,Bechtle:2020pkv,Bahl:2022igd}).
	\item[\textbullet]{\bf Flavor}:  Flavor constraints are also implemented in our analysis. We have used $B$-physics results, derived in \cite{Ouazghour:2023eqr}, as well as  the experimental data at 2$\sigma$ CL \cite{HeavyFlavorAveragingGroupHFLAV:2024ctg}  displayed in Table \ref{Tab2}.
\end{itemize}
{\renewcommand{\arraystretch}{1.5}
	{\setlength{\tabcolsep}{0.1cm} 
		\begin{table*}[t]
			\centering
			\setlength{\tabcolsep}{7pt}
			\renewcommand{\arraystretch}{1.2} %
			\begin{tabular}{|l||c|c|}
				\hline
				Observable & Experimental result & 95\% CL\\\hline
				BR($\bar{B}\to X_{s}\gamma$) \cite{Ouazghour:2023eqr}&$(3.49\pm 0.19)\times10^{-4}$ \cite{HeavyFlavorAveragingGroupHFLAV:2024ctg}&$[3.11\times 10^{-4} , 3.87\times 10^{-4}]$\\\hline
			\end{tabular}
			\caption{Experimental result on the flavor observable $\bar{B}\to X_{s}\gamma$ at 95$\%$ CL.}
			\label{Tab2}
		\end{table*}
Constraints specific to the parameter space of the 2HDMcT are as follows :
\begin{itemize}
\item \textbf{$\rho$ Parameter}:  
The presence of a scalar triplet modifies the $\rho$ parameter as  
\begin{eqnarray}
	\rho = \frac{v_0^2 + 2v_t^2}{v_0^2 + 4v_t^2} 
	\approx 1 - 2 \frac{v_t^2}{v_0^2} = 1 + \delta\rho.
	\label{eq:rho-thdmt}
\end{eqnarray}
In the SM, one has $\rho = 1$ at tree-level. Any departure from this value is tightly constrained and the latest global fit to EWPOs~\cite{ParticleDataGroup:2024cfk} yield  
\begin{eqnarray}
	\rho = 1.00031 \pm 0.00019 \, ,
\end{eqnarray}
which is about $1.6\sigma$ above the SM tree-level prediction.  
This small but nonzero deviation imposes an upper bound on the triplet VEV ($v_t$) in the 2HDMcT framework~\cite{Padhan:2019jlc,Ashanujjaman:2021txz}.  

\item \textbf{Lepton Flavor Violation (LFV)}:
From the Yukawa interaction shown in Eq.\,\ref{neutrino_mass}, {LFV} decays like $\mu\rightarrow e\gamma$ at loop-level and $\mu\rightarrow 3e$ at tree-level can be possible. The Branching Ratios (BRs) in the 2HDMcT can be calculated as \cite{Lavoura:2003xp,Kuno:1999jp,Akeroyd:2009nu,Ouazghour:2018mld} 
\begin{equation}
	{\rm BR}(\mu \rightarrow e \gamma) = 384 \pi^2 \, |A_R|^2
\end{equation}

\begin{equation}
	A_R = \frac{- q_e |(h^\dagger h)_{e\mu}|}{48 \sqrt{2} \pi^2 G_F}
	\left(
	\frac{C_{23}}{m_{H_1^\pm}^2}
	+ \frac{C_{33}}{m_{H_2^\pm}^2}
	+ \frac{8}{m_{H^{\pm\pm}}^2}
	\right)
\end{equation}

\begin{equation}
	{\rm BR}(\mu \rightarrow 3e)
	= \frac{|h_{ee}|^2 \, |h_{\mu e}|^2}
	{4 G_F^2 m_{H^{\pm\pm}}^4}
\end{equation}

with
\begin{equation}
	\begin{aligned}
		h_{ij}
		&= \frac{m_{ij}}{\sqrt{2} v_t} \\
		&= \frac{1}{\sqrt{2} v_t}
		\left(
		U_{\mathrm{PMNS}}
		\, \mathrm{diag}(m_1, m_2, m_3) \,
		U_{\mathrm{PMNS}}^T
		\right)_{ij}.
	\end{aligned}
\end{equation}

Here, $\alpha$ denotes the electromagnetic fine-structure constant and $G_F$ the Fermi constant, while $C_{23}$ and $C_{33}$ are elements of the charged-scalar rotation matrix. The explicit expressions for $C_{23}$ and $C_{33}$, as well as for the other rotation matrix elements, can be found in Ref.~\cite{Ouazghour:2018mld}. The upper bounds on the BRs of the above processes are $1.5 \times 10^{-13}$ for $\mu\rightarrow e\gamma$ \cite{MEGII:2025gzr} and $1.0 \times 10^{-12}$ for $\mu\rightarrow 3e$ \cite{SINDRUM:1987nra}. 
\item[\textbullet]{\bf Neutrino Oscillation Experiments}: 
The experimental constraints on the neutrino oscillation parameters~\cite{Esteban:2024eli} :  

\begin{align}
	& \ \Delta m_{21}^2= \ 7.49_{-0.19}^{+0.19} \times 10^{-5}\ \text{eV}^2, \\
	& \ \sin^2\theta_{12}= \ 0.308_{-0.011}^{+0.012} , \\
	& \sin^2\theta_{13} : 
	\begin{cases}
		\sin^2\theta_{13}= \ 0.02215_{-0.00058}^{+0.00056} \  & \text{(NH)}, \\[4pt]
		\sin^2\theta_{13}= \ 0.02231_{-0.00056}^{+0.00056} \  & \text{(IH)},
	\end{cases} \\[8pt]
	& \sin^2\theta_{23} : 
	\begin{cases}
		\sin^2\theta_{23} = \ 0.470_{-0.013}^{+0.017}  & \text{(NH)}, \\[4pt]
		\sin^2\theta_{23} = \ 0.550_{-0.015}^{+0.012} & \text{(IH)},
	\end{cases} \\[8pt]
	& |\Delta m_{3l}^2| : 
	\begin{cases}
		\ \Delta m_{3l}^2 = \ 2.513_{-0.019}^{+0.021}\times 10^{-3} \ \text{eV}^2 \ \ \text{(NH)}, \\[4pt]
		\ \Delta m_{3l}^2 = \ -2.484_{-0.020}^{+0.020}\times 10^{-3} \ \text{eV}^2 \ \ \text{(IH)},
	\end{cases}
\end{align}
Also cosmological observations requires the sum of neutrino masses to be~\cite{ParticleDataGroup:2024cfk},
\begin{equation}
\sum_{\nu=1}^3 m_{\nu} \ < 0.12\ \text{eV}.
\end{equation}
\end{itemize}

\begin{figure*}[tb]
\includegraphics[scale=0.4]{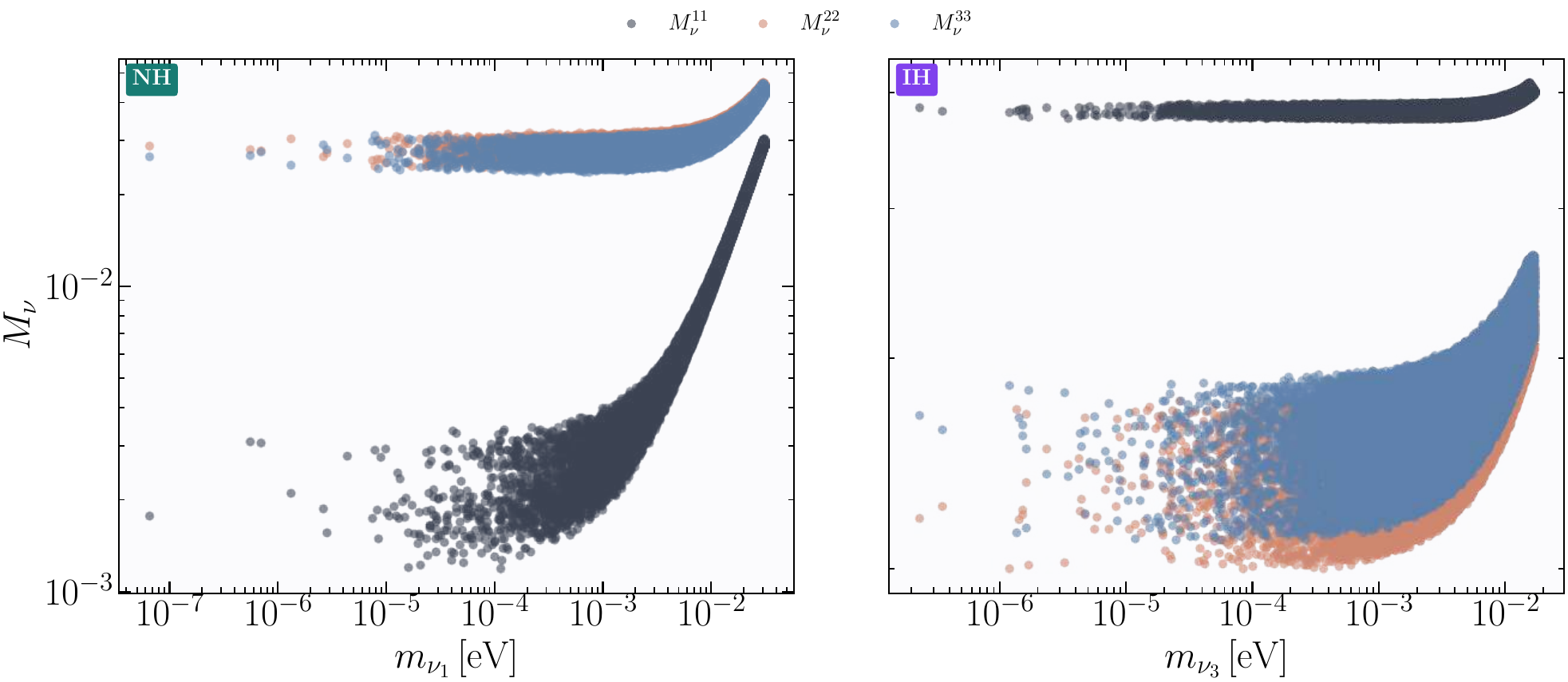}
\caption{Constraints on the diagonal entries of the neutrino mass matrix $M_\nu$ as functions of the lightest neutrino mass, $m_{\nu_1}$ for the NH (left) and $m_{\nu_3}$ for the IH (right). The phases are fixed to $\Phi_1 = 0$ and $\Phi_2 = 0$.
}
	\label{diagonal_elements}
\end{figure*}
\begin{figure*}[tb]
	\includegraphics[scale=0.4]{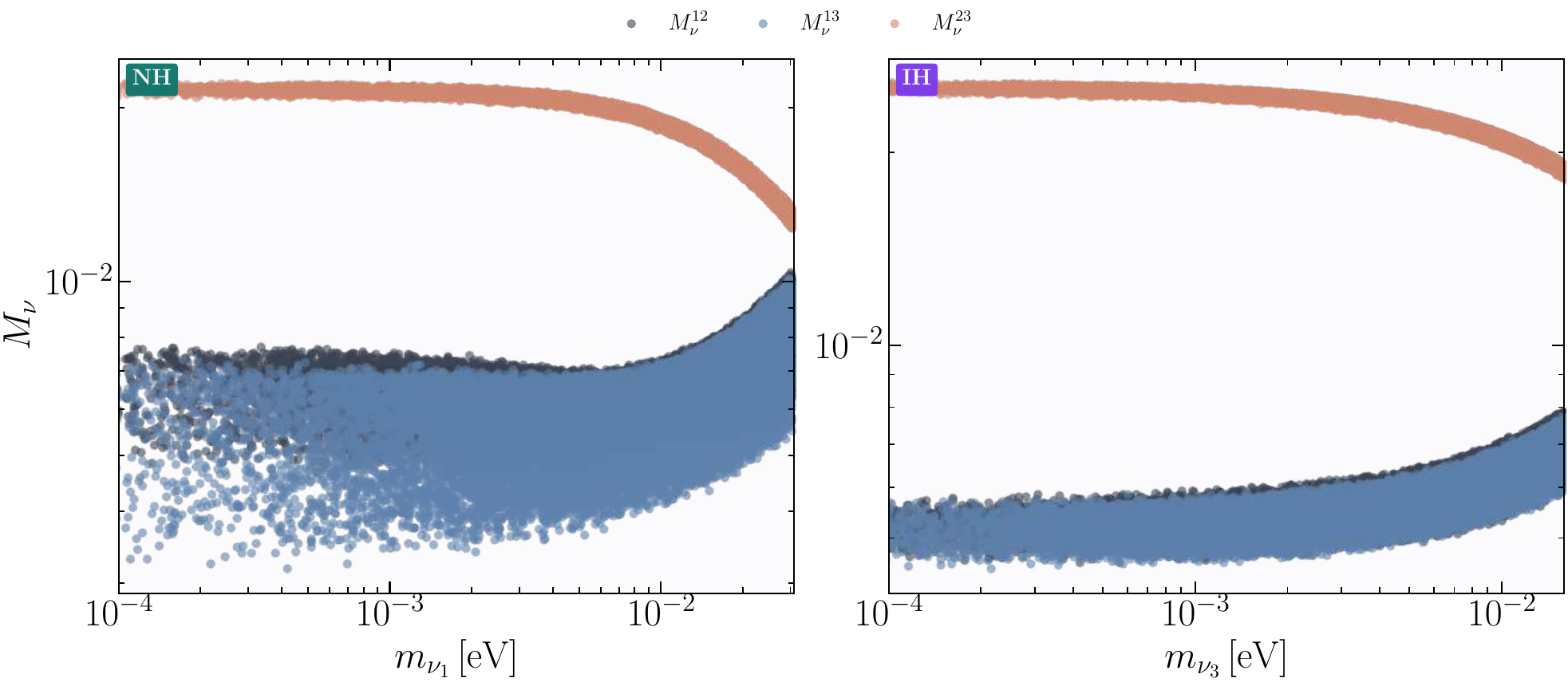}
	\caption{Constraints on the off-diagonal elements entries of the neutrino mass matrix $M_\nu$ as functions of the lightest neutrino mass, $m_{\nu_1}$ for the NH (left) and $m_{\nu_3}$ for the IH (right). The phases are fixed to $\Phi_1 = 0$ and $\Phi_2 = 0$.
	}
	\label{off-diagonal_elements}
\end{figure*}

The above experimental constraints, affect the allowed values for the neutrino mass matrix  $m_{\nu}$ as seen in Figs.~\ref{diagonal_elements} and \ref{off-diagonal_elements}, as a function of the lightest neutrino mass. For illustration we fix the phases to $\Phi_1 = 0$ and $\Phi_2 = 0$. In fact, these results directly reflect the structure of the neutrino mass matrix and the observed mixing pattern. In the case NH (left panel), the diagonal elements satisfy
\begin{equation}
	M_{\nu}^{11} < M_{\nu}^{22},\, M_{\nu}^{33},
\end{equation}
whereas for IH (right panel) one finds
\begin{equation}
	M_{\nu}^{11} > M_{\nu}^{22},\, M_{\nu}^{33}.
\end{equation}
Among the off-diagonal elements, $M_{\nu}^{23}$ attains the largest values in both hierarchies, reflecting the large atmospheric mixing angle, as illustrated in Fig.~\ref{off-diagonal_elements}.

\vspace{6pt}
\section{COMPUTATIONAL PROCEDURE and RESULTS}
\label{COMPUTATIONAL_PROCEDURE}

To identify the phenomenologically viable regions of the 2HDMcT we scan its parameter space while enforcing all theoretical and experimental constraints introduced earlier. 
The full set of independent input parameters considered in our analysis is 
\begin{align} \mathcal{P}_I = \big\{ &\alpha_1, \alpha_2, \alpha_3, m_{h_1}, m_{h_2}, \lambda_1, \lambda_3, \lambda_4, \lambda_6, \lambda_7, \lambda_8, \lambda_9, \bar{\lambda}_8, \bar{\lambda}_9,\nonumber\\ & \mu_1, v_t, \tan\beta \big\}. \label{parameters} 
\end{align}
These parameters are varied within the following ranges :
\begin{align}
	\begin{matrix}
		m_{h_1} = 125.09~\text{GeV},\qquad 
		m_{h_1} \leq m_{h_2} \leq m_{h_3} \leq 1000~\text{GeV}, \\[4pt]
		-\dfrac{\pi}{2} \leq \alpha_1 \leq \dfrac{\pi}{2},\qquad 
		-0.1 \leq \alpha_{2,3} \leq 0.1, \\[4pt]
		0.5 \leq \tan\beta \leq 120,\qquad 
		-10^2 \leq \mu_1 \leq 10^2, \\[4pt]
		0 \leq v_t \leq 2~\text{GeV},\qquad 
		-8\pi \leq \lambda_i,\, \bar{\lambda}_i \leq 8\pi,
	\end{matrix}
	\label{input}
\end{align}
with $\tan\beta = v_2 / v_1$.  Hereafter, 
the lightest CP-even state $h_1$ is taken to represent the observed Higgs boson, and its mass is fixed to $125$~GeV~\cite{ATLAS:2012yve,CMS:2012qbp}. Each parameter point generated during the scan is subjected to all relevant constraints and  
only the parameter points that satisfy these are subsequently passed to \texttt{FormCalc}~\cite{Hahn:2001rv,Hahn:1998yk,Kublbeck:1990xc}, which we employ to compute the cross-sections of the production channels considered in this work.
This procedure ensures that all reported results correspond exclusively to parameter configurations that are theoretically consistent and experimentally allowed.

As intimated, here, we investigate the electron-positron production of a doubly charged Higgs boson in association with a pair of singly charged Higgs bosons ($e^+ e^- \to H^{\pm\pm} H_1^\mp H_1^{\mp}$) as well as in association with a singly charged Higgs boson and a $W^\pm$ boson ($e^+ e^- \to H^{\pm\pm} H_1^\mp W^{\mp}$). The corresponding Feynman diagrams are displayed in Figs.~\ref{Diagram_HppHm1Hm1} and Fig.~\ref{Diagram_HppHp1W}, respectively.

For the process $e^+ e^- \to H^{\pm\pm} H_1^\mp H_1^{\mp}$, the amplitude receives its dominant contributions from $s$-channel topologies.  
The $s$-channel diagrams (Fig.~\ref{Diagram_HppHm1Hm1}$(d_{1\ldots 7})$) are mediated by a $\gamma$ or a $Z$ boson.  

For the associated production channel $e^+ e^- \to H^{\pm\pm} H_1^\mp W^{\mp}$, the amplitude arises from a broader set of gauge and Higgs  interactions, as illustrated in Fig.~\ref{Diagram_HppHp1W}.  
in this case, the diagrams may be grouped into $s$-channel and $t$-channel classes. Specifically,  the $s$-channel contributions (Fig.~\ref{Diagram_HppHp1W}$(d_{1\ldots 7})$) proceed through the exchange of a $\gamma$, or a $Z$ boson.  
These include configurations where the intermediate vector boson subsequently splits into a $W^\pm H_1^\mp$.  
The $t$-channel contribution (Fig.~\ref{Diagram_HppHp1W}$(d_{8})$) arise from neutrino exchange and are characteristic of processes that produce a final-state $W^\pm$ boson, which emission can take place either from an  internal neutrino  or  Higgs propagator.

\subsection*{Relevant Couplings}

The production mechanisms for $e^+ e^- \to H^{\pm\pm} H_1^\mp H_1^{\mp}$ and  
$e^+ e^- \to H^{\pm\pm} H_1^\mp W^{\mp}$ are controlled by a set of gauge and Higgs  interactions characteristic of the 2HDMcT framework.  
Here, we briefly summarize the most important couplings that govern the size and behavior of the amplitudes.

\paragraph{Gauge Interactions.}  
The doubly charged Higgs boson couples  to the EW gauge bosons through its $SU(2)_L$ triplet nature.  
The most relevant vertices are as follows.
\begin{itemize}
\item $H^{++}H^{--}\gamma$ and $H^{++}H^{--}Z$:  
These arise from covariant-derivative interactions (see Eqs.~\ref{eq:covd1} and \ref{eq:covd2}) and play a central role in the $s$-channel production via photon and $Z$ exchange.
\item $H^{++}H^-_1 W^-$:  
A gauge interaction proportional to the EW coupling $g$.  
This vertex contributes to both the $e^-e^+ \to H^{\pm\pm} H_1^\mp H_1^{\mp}$ and $e^-e^+ \to H^{\pm\pm} H_1^\mp W^{\mp}$ processes.
\item $W^- \nu_e e^- $:  
The charged-current interaction, responsible for neutrino-mediated $t$-channel diagrams in both processes. This vertex is essential for the process $e^-e^+ \to H^{\pm\pm} H_1^\mp W^{\mp}$.
\end{itemize}

\paragraph{Higgs Interactions.}  
The extended Higgs  sector of the 2HDMcT gives rise to multiple trilinear and quartic  interactions involving both charged and neutral Higgs states (see the Higgs potential in Eq.~\ref{scalar_pot}). Both processes are directly sensitive to the detail of the scalar potential through diagrams like $d_2$ and others. The most relevant interactions are listed below. 
\begin{itemize}
	\item $H^{--}H^-_1H^-_1$:  
	A trilinear scalar coupling generated by the triplet–doublet mixing. This vertex contributes to both the $e^-e^+ \to H^{\pm\pm} H_1^\mp H_1^{\mp}$ and $e^-e^+ \to H^{\pm\pm} H_1^\mp W^{\mp}$ processes.
	
	\item $H^{++}H^+_2H^-_1$:  
	A trilinear scalar coupling generated by the triplet–doublet mixing. It appears in diagrams mediated by a $Z$ boson, such as Fig.~\ref{Diagram_HppHm1Hm1}$(d_{6})$ and Fig.~\ref{Diagram_HppHp1W}$(d_{2})$. This vertex also enters both the $e^-e^+ \to H^{\pm\pm} H_1^\mp H_1^{\mp}$ and $e^-e^+ \to H^{\pm\pm} H_1^\mp W^{\mp}$ processes.
	
	\item $h_k H^{++} H^{--}$ and $h_k H^-_i H^+_j$:  
	Couplings of the neutral CP-even Higgs bosons ($k = 1,2,3$) to singly charged ($i,j=1,2$) and doubly charged Higgs states.  
	These contribute to the $s$-channel diagrams, which are suppressed due to the small electron Yukawa coupling.
\end{itemize}

\medskip

Overall, the competition between gauge and scalar interactions sets the diagram hierarchy and dictates the cross-section behavior within the allowed 2HDMcT parameter space.
\subsection{Cross-sections}

As previously mentioned, the mass splitting between the doubly charged Higgs state and the singly charged ones plays a role in evading EWPO constraints in the 2HDMcT, hence, we present it in Fig.~\ref{mHpp_mHp1}. From here, it is clear that the mass difference $m_{H^{++}}-m_{H^+_1}$ can be quite significant whereas $m_{H^{++}}$ and $ m_{H^+_2}$ are nearly degenerate. In fact, such a mass gap is also important in driving a hierarchy between the production rates of $e^+e^-\to H^{\pm\pm} H_1^\mp H_1^{\mp}$ relative to those for $e^+e^-\to H^{++}H^{--}$ followed by $H^{\pm\pm}\to H^\pm_1H^\pm_1$ decays (and consequently also in the case of $H^{\pm\pm}\to H^\pm_1W^\pm$ transitions).

In Fig.~\ref{results_HppHm1Hm1}, we present the cross-section $\sigma$ for the process 
$e^+ e^- \to H^{\pm\pm} H_1^\mp H_1^{\mp}$ at $\sqrt{s} = 500$, $1000$, and $1500~\text{GeV}$.  
This production channel is primarily governed by the trilinear scalar vertex $H^{++}H_1^-H_1^-$ and the gauge-Higgs vertex $H^{++}H_1^-W^-$, which are proportional to the couplings $\lambda_{H^{++}H_1^-H_1^-}$ and $C_{23}$, respectively.  
As a result, the majority of the contributing diagrams depend sensitively on these interactions.  
The clear message stemming from this figure is that the $H^{\pm\pm} H_1^\mp H_1^{\mp}$ final state is largely driven by $H^{++}H^{--}$ production (the empty red diamonds curves) followed by $H^{\pm\pm}\to H_1^\pm H_1^\pm$, yet, the contribution from all other  diagrams can be significant not just when the latter decay is forbidden ($m_{H^{--}}<2m_{H_1^-}$) but also when it is not, 
thus pointing to their significance (in part also aided by off-shellness effect in such a decay), in fact, raising the    
$\sigma(e^+ e^- \to H^{++}H^{--})\times {\rm BR}(H^{\pm\pm}\to H_1^\pm H_1^\pm)$ rates by more than a factor of 2 in the $m_{H^{--}}>2m_{H_1^-}$ region. Such an effect is visible at all energies considered, altogether yielding excess cross-sections (i.e., well beyond those from on-shell pair production and decay) already for $m_{H^{++}}$ values around 180 GeV.

In a similar manner, Fig.~\ref{results_HppHm1W} displays the cross-sections for the process  
$\sigma(e^+ e^- \to H^{\pm\pm} H_1^\mp W^{\mp})$ at $\sqrt{s} = 500$, $1000$, and $1500~\text{GeV}$.  
This channel is predominantly controlled by the trilinear scalar interaction $H^{++}H_1^-H_1^-$ and the gauge--scalar interaction $H^{++}H_1^-W^-$, which are governed by the couplings $\lambda_{H^{++}H_1^-H_1^-}$ and $C_{23}$, respectively. Only the diagrams $(d_1)$ and $(d_{7})$ are independent of these two vertices.
Consequently, the majority of the diagrams contributing to this process exhibit pronounced sensitivity to the values of these two couplings. The cross sections arising from the $2\to3$ production mode can again exceed those obtained from the $2\to2$ production followed by decay, however, this occurs only for significantly larger values and within a narrower mass range of the doubly charged Higgs boson (and only at $\sqrt{s}=1000$ and $1500~\text{GeV}$).

Therefore, we define four Benchmark Points (BPs) for which the discussed $2\to 3$ body cross-section dominates the $2\to2$ body one (times BR), as exemplified by the green stars in the center and right panels of Figs.~\ref{results_HppHm1Hm1} and \ref{results_HppHm1W}. The corresponding input parameter values are given in Tab.~\ref{BPs}. We will use these in the forthcoming MC analysis, which we will limit to the CoM energies of 1000 and 1500 GeV.

\begin{figure*}[!t]  
	\centering
	\includegraphics[scale=0.75]{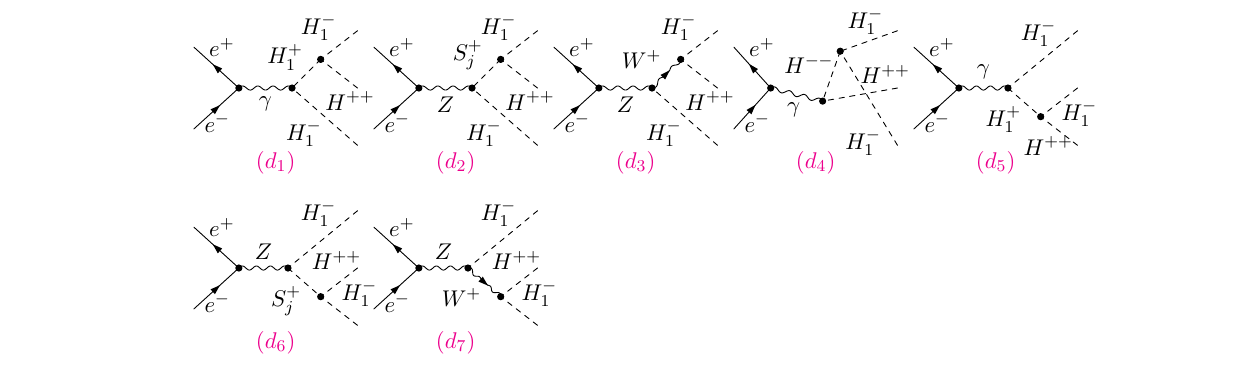}
	\\
	\caption{
		\small
		Tree-level generic Feynman diagrams for $e^-e^+\to H^{++}H_1^-H_1^-$ are shown in $(d_{1,...,7})$. For all diagrams $S_j^\pm$ refers to $H_1^{\pm}$ or $H_2^{\pm}$.}
	\label{Diagram_HppHm1Hm1}
\end{figure*}
\begin{figure*}[!htb]  
	\centering
	\includegraphics[scale=0.75]{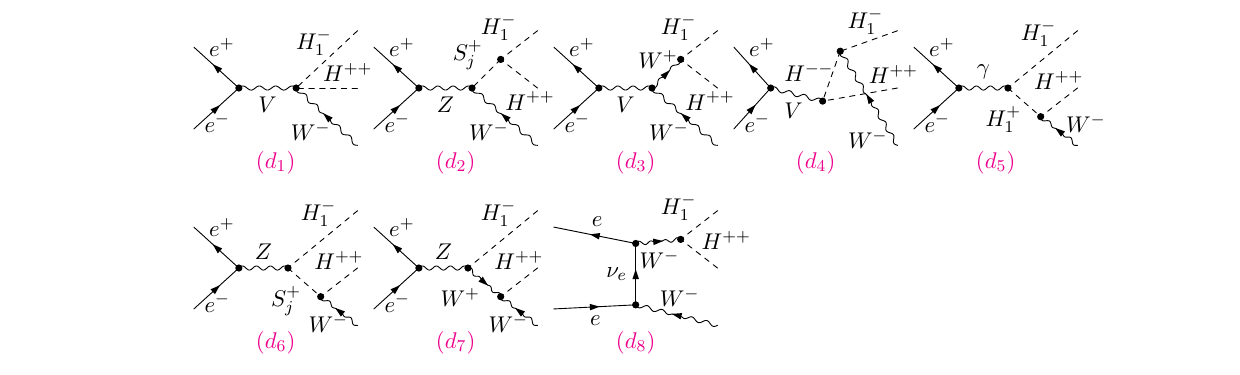}
	\\
	\caption{
		\small
		Tree-level generic Feynman diagrams for $e^-e^+\to H^{++}H_1^-W^-$ are shown in $(d_{1,...,8})$. For all diagrams $S_j^\pm$ refers to $H_1^{\pm}$ or $H_2^{\pm}$. Furthermore, $V=\gamma,Z$.}
	\label{Diagram_HppHp1W}
\end{figure*}
\begin{figure*}[!htb]
	\centering

	\includegraphics[scale=0.4]{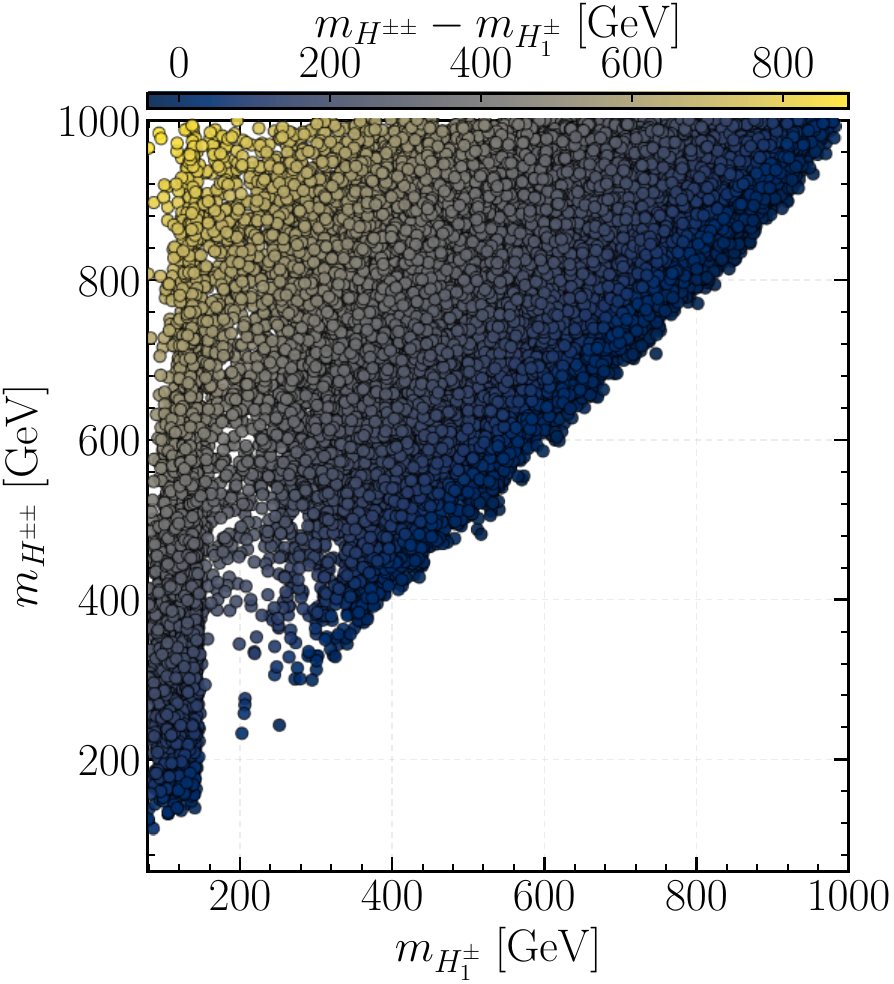}
	\includegraphics[scale=0.4]{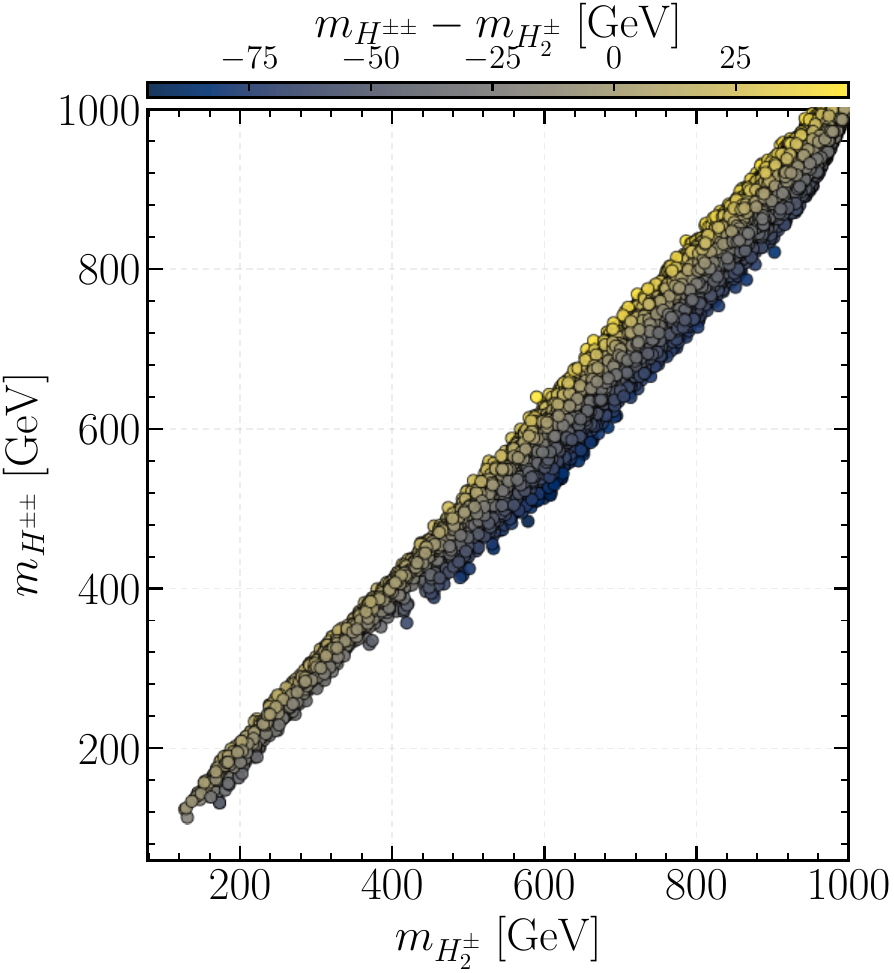}
	\\
\caption{
	\small
	Mass splitting between $H^{\pm\pm}$ and $H_1^{\pm}$ (left), and between $H^{\pm\pm}$ and $H_2^{\pm}$ (right).
}
	\label{mHpp_mHp1}
\end{figure*}
\begin{figure*}[!h]
	\centering
	\includegraphics[scale=0.33]{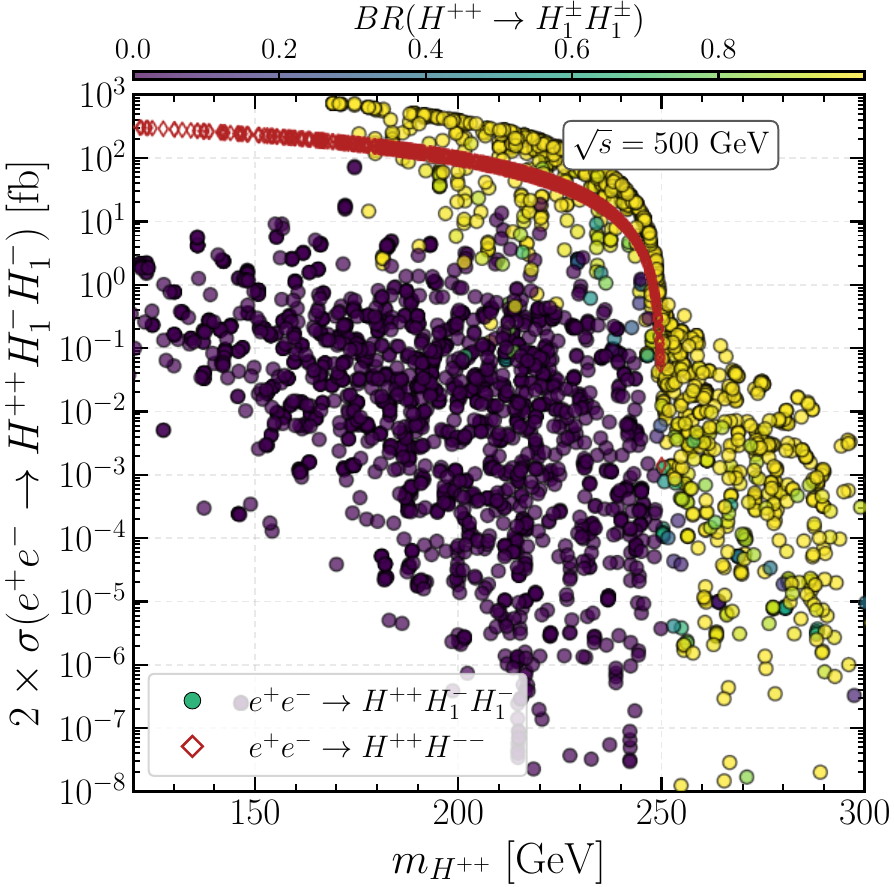}
	\includegraphics[scale=0.33]{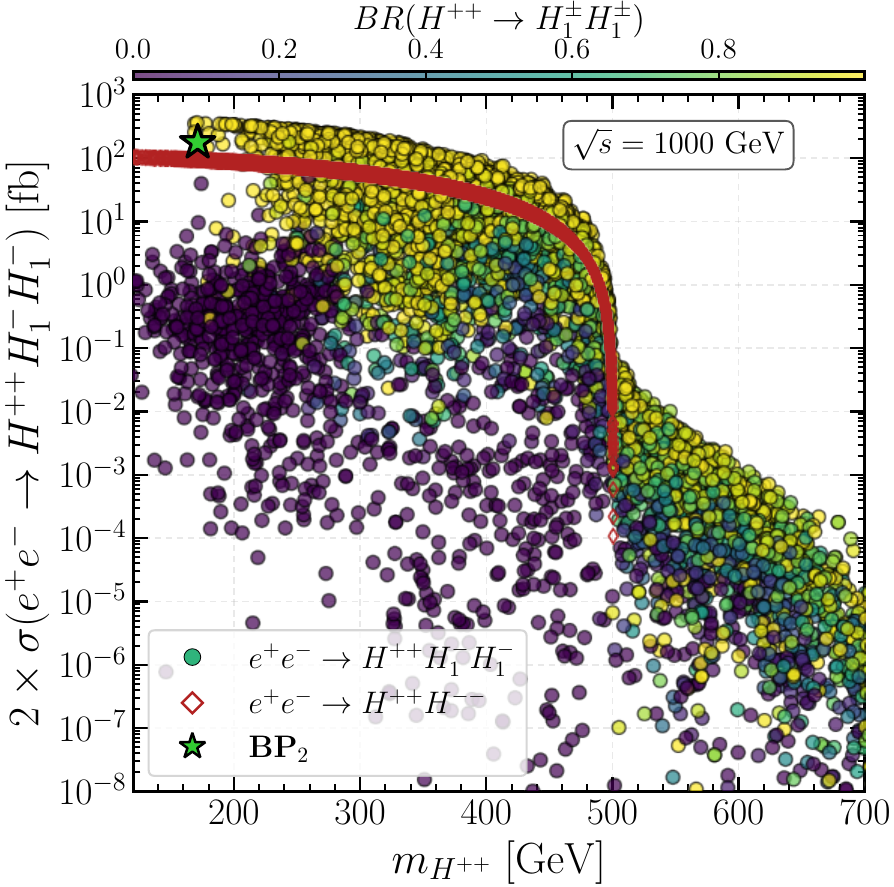}
	\includegraphics[scale=0.33]{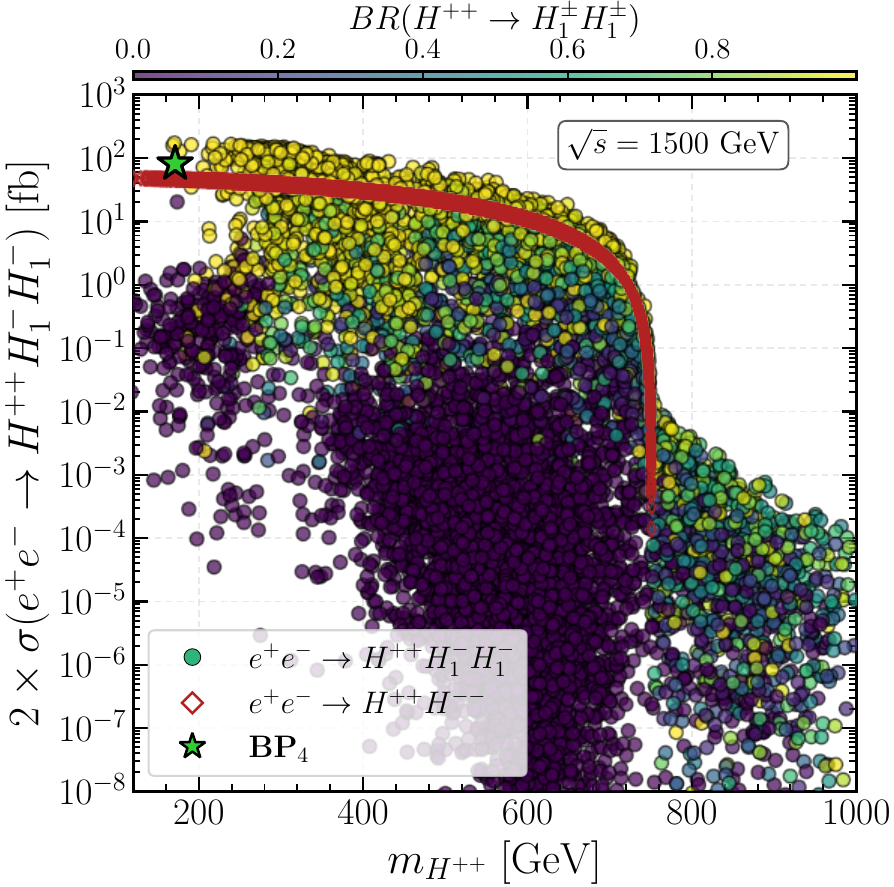}
	\\
	\caption{
		\small
		Production cross-sections for $e^+e^- \to H^{\pm\pm} H_1^{\mp} H_1^{\mp}$ as a function of $m_{H^{++}}$ at $\sqrt{s}=500,\ 1000,\ 1500$ GeV. }
	\label{results_HppHm1Hm1}
\end{figure*}	
\begin{figure*}[!htb]
	\centering
	\includegraphics[scale=0.33]{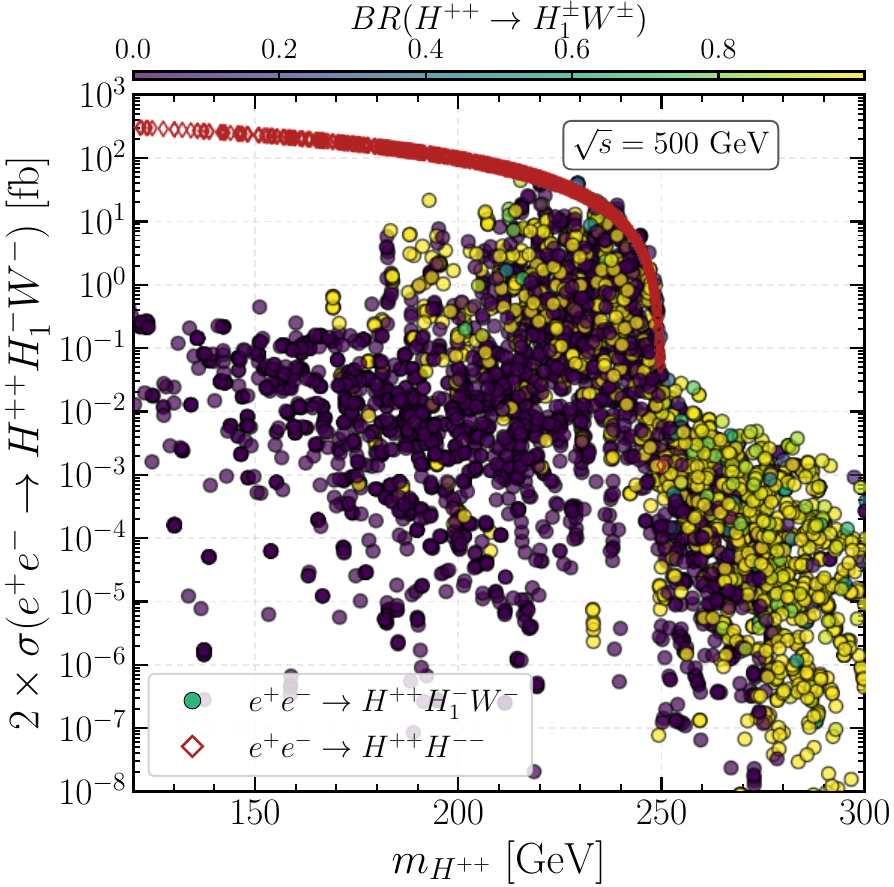}
	\includegraphics[scale=0.33]{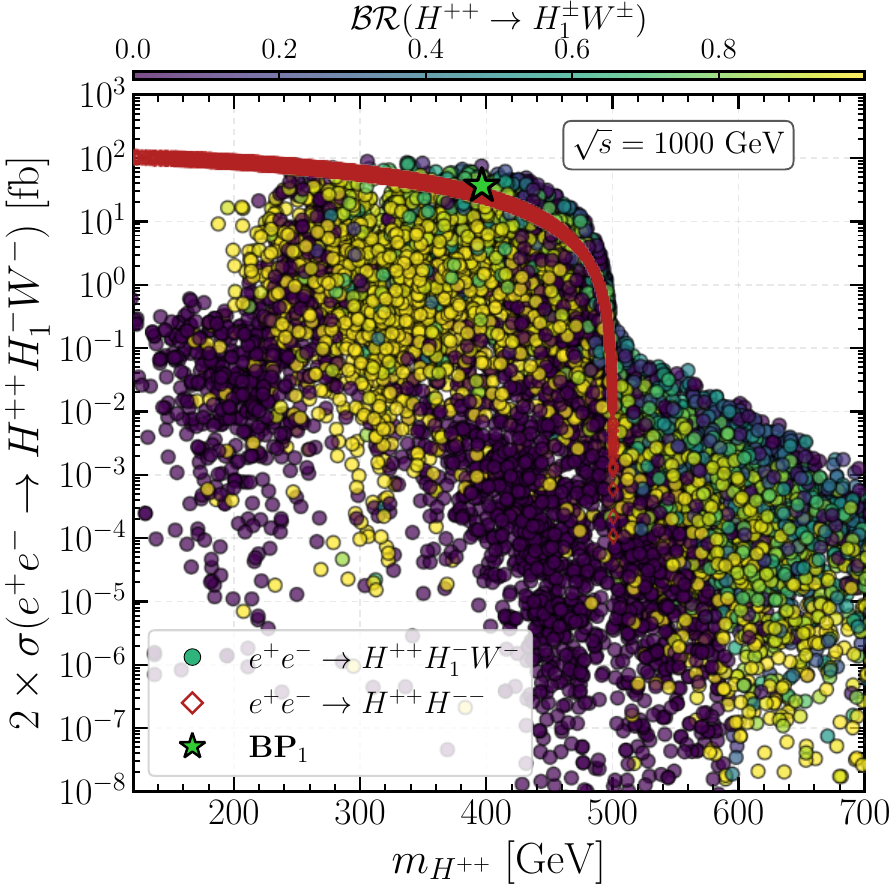}
	\includegraphics[scale=0.33]{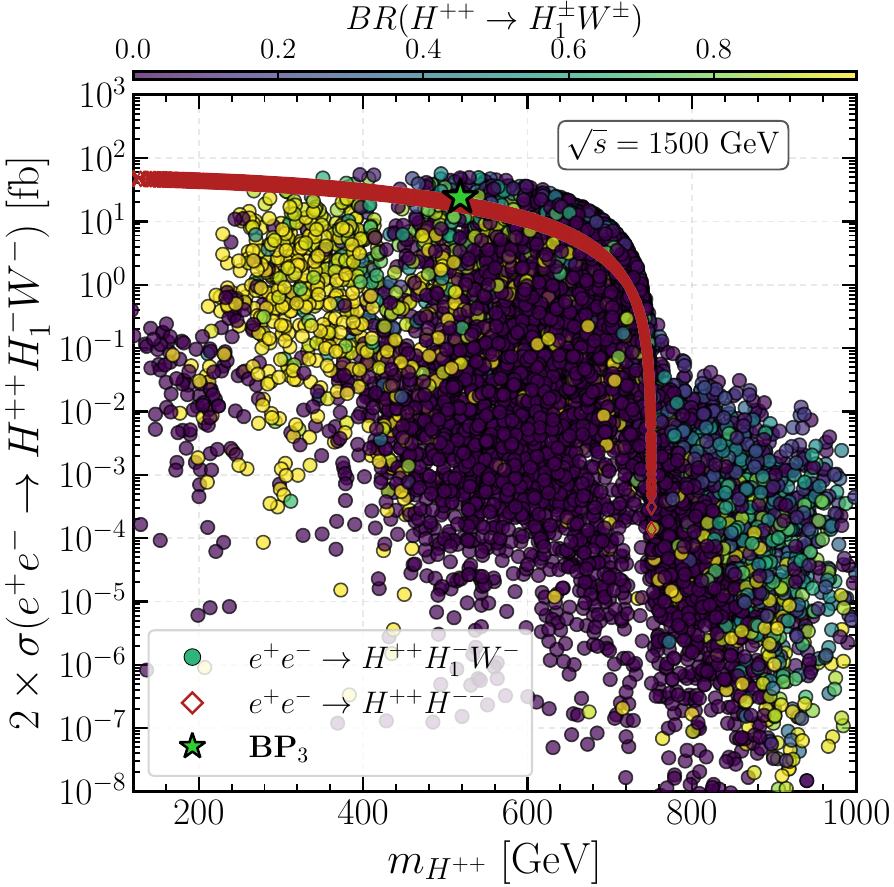}
	\\
	\caption{
		\small
		Production cross-sections for $e^+e^- \to H^{\pm\pm} H_1^{\mp} W^{\mp}$ as a function of $m_{H^{++}}$ at $\sqrt{s}=500,\ 1000,\ 1500$ GeV. }
	\label{results_HppHm1W}
\end{figure*}
\begin{table*}[!h]
	\centering
	\setlength{\tabcolsep}{1.5pt}
	\renewcommand{\arraystretch}{1.1}
	\resizebox{\textwidth}{!}{%
		\begin{tabular}{|c|c|c|c|c|c|c|c|c|c|c|c|c|c|c|c|c|c|}
			\hline\hline
			& $m_{h_1}$ & $m_{h_2}$ & $\lambda_1$ & $\lambda_3$ & $\lambda_4$ & $\lambda_6$ & $\lambda_7$ & $\lambda_8$ & $\lambda_9$ & $\bar{\lambda}_8$ & $\bar{\lambda}_9$ & $\tan\beta$ & $\alpha_1$ & $\alpha_2$ & $\alpha_3$ & $v_t$ & $\sigma\,(\mathrm{fb})$ \\
			\hline
			BP1 & 125.09 & 164.40 & 0.7043 & $-6.22\!\times\!10^{-2}$ & 0.5547 & 0.3290 & 0.0862 & $-0.1013$ & 0.2678 & 0.2703 & 0.8361 & 118.71 & $-1.5636$ & $-3.68\!\times\!10^{-4}$ & $-4.28\!\times\!10^{-2}$ & 0.4271 &  36.58\\
			\hline
			BP2 & 125.09 & 145.30 & 0.5678 & $-4.03\!\times\!10^{-2}$ & 0.2655 & 0.7227 & $7.99\!\times\!10^{-2}$ & $-0.5346$ & $7.68\!\times\!10^{-2}$ & 2.0904 & 0.3526 & 82.41 & $-1.5615$ & $-2.96\!\times\!10^{-4}$ & $-3.30\!\times\!10^{-2}$ & $5.88\!\times\!10^{-2}$ &177.02  \\
			\hline
			BP3 & 125.09 & 266.81 & 0.1888 & 1.9616 & $-2.74\!\times\!10^{-2}$ & 7.0303 & 0.9439 & $-8.54\!\times\!10^{-2}$ & 0.9211 & $-0.7390$ & 1.8138 & 115.93 & $-1.5628$ & $-1.40\!\times\!10^{-4}$ & $-6.13\!\times\!10^{-2}$ & 0.4143 &  24.00 \\
			\hline
			BP4 & 125.09 & 145.30 & 0.5678 & $-4.03\!\times\!10^{-2}$ & 0.2655 & 0.7227 & 0.0799 & $-0.5346$ & 0.0768 & 2.0904 & 0.3526 & 92.58 & $-1.5615$ & $-2.96\!\times\!10^{-4}$ & $-3.30\!\times\!10^{-2}$ & 0.0588 & 81.83 \\
			\hline\hline
		\end{tabular}
	}
	\caption{BPs used for the signal-to-background analysis.}
	\label{BPs}
\end{table*}

\subsection{Signal-to-background Analysis}
\label{sec:SB}

To quantify the observability of the doubly charged Higgs boson $H^{\pm\pm}$ predicted in the 2HDMcT with the type-X Yukawa texture\footnote{Note that the production cross sections are largely independent of the Yukawa texture, as they are governed primarily by gauge interactions; the dependence on the type-X structure enters through the decays of the singly charged Higgs states.}, crucially, in the region dominated by the discussed $2\to 3$ body processes, we develop search strategies aimed at maximizing the separation between the ensuing signals and  SM backgrounds. The analysis is based on a state-of-the-art simulation chain that accounts for matrix-element generation, resonance decays, parton showering, hadronization, heavy flavor decays, jet reconstruction, and detector effects. As a representative case study, we focus on the clean multilepton signature $ \ell^+\ell^+\ell^-\ell^-+\slashed{E}_T $ (henceforth, $4\ell+\slashed{E}_T$ for short), which we will prove to provide strong signal sensitivity once the dominant backgrounds are efficiently suppressed. For each collider energy considered in this work, we select a representative BP from the scan, reported in Tab.~\ref{BPs}. The production cross-sections discussed above indicate that the signal rates are potentially accessible, provided that a dedicated background rejection strategy is implemented.

\paragraph{MC Event Generation and Detector Simulation.}
Signal events are generated at parton level with \texttt{MadGraph5\_aMC\_v3.4.2}~\cite{Alwall:2014hca,Hagiwara:2012vz}, interfaced to \texttt{Pythia-8.20}~\cite{Sjostrand:2007gs} for parton showering,  hadronization, and heavy flavor decays. Jets are clustered using \texttt{FastJet}~\cite{Cacciari:2011ma}. Detector effects are modelled with \texttt{Delphes-3.4.5}~\cite{deFavereau:2013fsa}, using the default \texttt{Delphes\_Card\_ILCgen}\footnote{https://github.com/iLCSoft/ILCDelphes.} detector card, which implements the anti-$k_t$ algorithm~\cite{Cacciari:2008gp}. Finally, both the signals and background events are analyzed with \texttt{MadAnalysis5} \cite{Conte:2012fm, Conte:2013mea}.

\paragraph{Signal Channels and Decay Topologies.}
The $4\ell+\slashed{E}_T$ signature is obtained from two different production and decay patterns. For the first channel, we consider $H^{\pm\pm} H_1^\mp W^{\mp}$ production followed by the cascade $H^{\pm\pm}\to H_1^\pm W^\pm$ and $H_1^\pm \to \tau_\ell \nu_\tau$ with leptonic $W^\pm$ decays:
\begin{eqnarray}
	e^+ e^- \to H^{\pm\pm} H_1^\mp W^{\mp}  &\to& H_1^{\pm}W^{\pm}\,H_1^\mp W^{\mp}\nonumber\\
	&\to& \tau_\ell^+\,\ell^+\,\tau_\ell^-\,\ell^- +\slashed{E}_T \nonumber\\
	&\to& 4\ell +\slashed{E}_T \, . \nonumber
\end{eqnarray}
The second channel is $H^{\pm\pm} H_1^\mp H_1^{\mp}$ production with $H^{\pm\pm}\to H_1^\pm H_1^\pm$ and leptonic $\tau$ decays:
\begin{eqnarray}
	e^+ e^- \to H^{\pm\pm} H_1^\mp H_1^{\mp}  &\to& H_1^{\pm} H_1^{\pm} H_1^\mp H_1^{\mp} \nonumber\\
	&\to& \tau_\ell^+\tau_\ell^+\tau_\ell^- \tau_\ell^- +\slashed{E}_T \nonumber\\
	&\to& 4\ell +\slashed{E}_T \, . \nonumber
\end{eqnarray}
Here $\tau^\pm$ denotes the charged $\tau$ lepton, $\tau_\ell$ indicates its leptonic decay mode, and $\ell=e,\mu$ are light charged leptons. The missing transverse energy $\slashed{E}_T$ originates from neutrinos in $\tau$ and $W^\pm$ decays.

\paragraph{SM backgrounds.}
The dominant SM backgrounds to the $4\ell+\slashed{E}_T$ signature arise from multiboson and top-associated production:
$W^+W^-Z$, $ZZZ$, $ZZ\gamma$, $t\bar t Z$, $t\bar t\gamma$, and $W^+W^-W^+W^-$.
We assume semileptonic top decays and leptonic $W^\pm$ decays throughout. The $Z$ boson is taken to decay into $\ell^+\ell^-$ or $\nu_\ell\bar\nu_\ell$, while $\gamma\to \ell^+\ell^-$ is imposed when relevant. The background channels are generated according to
the following list.
\begin{itemize}
	\item $e^+ e^- \rightarrow W^- W^+ Z,\ (W^- \rightarrow \ell^- \bar{\nu}_\ell),\ (W^+ \rightarrow \ell^+ \nu_\ell),\ (Z \rightarrow \ell^+ \ell^-).$
	\item $e^+ e^- \rightarrow Z Z Z,\ (Z \rightarrow \ell^+ \ell^-),\ (Z \rightarrow \ell^+ \ell^-),\ (Z \rightarrow \nu_\ell \bar{\nu}_\ell).$
	\item $e^+ e^- \rightarrow Z Z \gamma,\ (Z \rightarrow \ell^+ \ell^-),\ (\gamma \rightarrow \ell^+ \ell^-),\ (Z \rightarrow \nu_\ell \bar{\nu}_\ell).$
	\item $e^+ e^- \rightarrow t\bar{t}Z,\ (t \rightarrow W^+ b,\ W^+ \rightarrow \ell^+ \nu_\ell),\ (\bar{t} \rightarrow W^- \bar{b},\ W^- \rightarrow \ell^- \bar{\nu}_\ell),\ (Z \rightarrow \ell^+ \ell^-).$
	\item $e^+ e^- \rightarrow t\bar{t}\gamma,\ (t \rightarrow W^+ b,\ W^+ \rightarrow \ell^+ \nu_\ell),\ (\bar{t} \rightarrow W^- \bar{b},\ W^- \rightarrow \ell^- \bar{\nu}_\ell),\ (\gamma \rightarrow \ell^+ \ell^-).$
	\item $e^+ e^- \rightarrow W^+ W^- W^+ W^-,\ (W^- \rightarrow \ell^- \bar{\nu}_\ell),\ (W^+ \rightarrow \ell^+ \nu_\ell),\ (W^- \rightarrow \ell^- \bar{\nu}_\ell),\ (W^+ \rightarrow \ell^+ \nu_\ell).$
\end{itemize}

\paragraph{Object Preselection.}
Events are required to satisfy the baseline acceptance and isolation criteria
\begin{equation}
	|\eta^{j,\ell}| < 2.5,\qquad \Delta R^{\ell\ell,\ell j,jj} \geq 0.5,
	\label{Bcuts}
\end{equation}
ensuring reconstructed leptons and jets remain within the fiducial region and are well separated.

\paragraph{Significance Definition.}
The signal sensitivity $\mathcal{Z}$  is quantified using the median discovery significance of Ref.~\cite{Cowan:2010js}. For signal and background yields $s$ and $b$, respectively, and a fractional systematic uncertainty $\delta$ on the background, we compute
\begin{widetext}
	\begin{align}
		\mathcal{Z} =
		\sqrt{2\Bigg[\left(s+b\right)\ln\!\left(\frac{\left(s+b\right)\left(1+\delta^2b\right)}{b+\delta^2b\left(s+b\right)}\right)
			-\frac{1}{\delta^2}\ln\!\left(1+\delta^2\frac{s}{1+\delta^2b}\right)\Bigg]}\, .
	\end{align}
\end{widetext}

\begin{itemize}
	\centering \item \bf{Processes at $\sqrt{s} = 1000~\rm{GeV}$}
\end{itemize}

We illustrate in Fig.~\ref{p1} the kinematic distributions for the $4\ell + \slashed{E}_T$ final state at $\sqrt{s}=1000$ GeV. These include the pseudorapidity $\eta(\ell_1)$ of the leading lepton (left panel), the transverse momentum $p_T(\ell_1)$ (middle panel), and the missing transverse energy $\slashed{E}_T$ (right panel).

\underbar{$e^{+} e^{-} \rightarrow H^{\pm\pm} H_1^\mp W^{\mp} \rightarrow  4\ell + \slashed{E}_T$ {\textcolor{red}{ (BP1)}}}
Guided by these distributions, we construct a sequential cut strategy summarized in Tab.~\ref{selection_cuts_HppHp1W_BP1}. We first suppress top-induced backgrounds by rejecting events with more than one $b$-tagged jet, $N(b)\leq 1$, which removes about $78\%$ of the $t\bar tZ/\gamma$ contributions. We then restrict the leading-lepton pseudorapidity to the central region, $-0.9<\eta(\ell_1)<0.9$, eliminating most of the remaining $t\bar tZ/\gamma$ events and about $50\%$ of $W^+W^-Z$, while retaining $\sim 87.5\%$ of the signal. An upper bound $P_T(\ell_1)<300$~GeV further reduces the $W^+W^-Z$ background by $64\%$ with only a modest loss in signal efficiency. Finally, requiring $\slashed{E}_T<300$~GeV yields an additional suppression of about $85\%$ of the top backgrounds and $65\%$ of $W^+W^-Z$. The cumulative impact of these requirements on the final  cross-sections is reported in Tab.~\ref{cutft_HppHp1W_BP1}.

\underbar{$e^{+} e^{-} \rightarrow H^{\pm\pm} H_1^\mp H_1^{\mp} \rightarrow  4\ell + \slashed{E}_T$  {\textcolor{red}{(BP2)}}}
An analogous optimization is performed for this channel, with the selection requirements listed in Tab.~\ref{selection_cuts_HppHp1Hp1_BP2}. After the $b$-jet requirement $N(b)\leq 1$, about $78\%$ of the $t\bar tZ$ and $t\bar t\gamma$ backgrounds are rejected. Constraining the leading-lepton pseudorapidity to $-1.6<\eta(\ell_1)<1.6$ suppresses the bulk of the top-induced contributions and about $18\%$ of $W^+W^-Z$, while retaining $97.6\%$ of the signal. A tighter transverse-momentum cut $P_T(\ell_1)<200$~GeV reduces the remaining $W^+W^-Z$ contribution by $60\%$ with a signal efficiency of $97.3\%$. The requirement $\slashed{E}_T<200$~GeV further improves rejection, removing roughly $86\%$ of $t\bar tZ/\gamma$ and $65\%$ of $W^+W^-Z$. The full cut-flow is shown in Tab.~\ref{cutft_HppHp1Hp1_BP2}.

\begin{itemize}
	\centering \item \bf{Processes at $\sqrt{s} = 1500~\rm{GeV}$}
\end{itemize}

The corresponding kinematic distributions at $\sqrt{s}=1500$~GeV are presented in Fig.~\ref{p2}. As before, the pseudorapidity $\eta(\ell_1)$ of the leading lepton (left panel), the transverse momentum $p_T(\ell_1)$ (middle panel), and the missing transverse energy $\slashed{E}_T$ (right panel).

\underbar{$e^{+} e^{-} \rightarrow H^{\pm\pm} H_1^\mp W^{\mp} \rightarrow  4\ell + \slashed{E}_T$  {\textcolor{red}{(BP3)}}}
Based on these distributions, we define the optimized selection summarized in Tab.~\ref{selection_cuts_HppHp1W_BP3}. The requirement $N(b)\leq 1$ removes about $80\%$ of the $t\bar tZ$ and $t\bar t\gamma$ backgrounds. Restricting the leading-lepton pseudorapidity to $-0.9<\eta(\ell_1)<0.9$ suppresses roughly $88\%$ of the remaining top backgrounds and about $59\%$ of $W^+W^-Z$, while maintaining $89.2\%$ of the signal. We then impose $P_T(\ell_1)<400$~GeV, which reduces the remaining EW  backgrounds substantially, followed by $\slashed{E}_T<400$~GeV, which eliminates about $90\%$ of the top-induced backgrounds and about $79\%$ of $W^+W^-Z$. The corresponding cut-flow is given in Tab.~\ref{cutft_HppHp1W_BP3}.

\underbar{$e^{+} e^{-} \rightarrow H^{\pm\pm} H_1^\mp H_1^{\mp} \rightarrow  4\ell + \slashed{E}_T$  {\textcolor{red}{(BP4)}}}
For this topology, the optimized selection strategy is summarized in Tab.~\ref{selection_cuts_HppHp1Hp1_BP4}. After applying $N(b)\leq 1$, nearly $79\%$ of the $t\bar tZ$ and $t\bar t\gamma$ backgrounds are removed. Requiring $-1.4<\eta(\ell_1)<1.4$ further suppresses the remaining top backgrounds and about $20\%$ of the $W^+W^-Z/\gamma$ contributions, while keeping the signal efficiency close to $94\%$. Imposing $P_T(\ell_1)<300$~GeV reduces the $W^+W^-Z$ background by $71\%$, with $93.7\%$ of the signal retained. Finally, $\slashed{E}_T<300$~GeV yields an additional rejection of about $90\%$ of $t\bar tZ/\gamma$ and $75\%$ of $W^+W^-Z$. The resulting cross-section evolution through the cut-flow is reported in Tab.~\ref{cutft_HppHp1Hp1_BP4}.
\begin{itemize}
	\centering \item \bf{ Summary of sensitivities}
\end{itemize}
In Fig.~\ref{Significance_AllBP} we present the discovery significance $\mathcal{Z}$ for all BPs, explicitly indicating the associated production processes. Systematic uncertainties of $\delta=5\%$ and $10\%$ are included for $\sqrt{s}=1000$ and $1500$~GeV, and the projected sensitivities are shown for integrated luminosities of $\mathcal{L}=500$, $1000$, and $1500~\mathrm{fb}^{-1}$. The results demonstrate that, already at $\mathcal{L}=500~\mathrm{fb}^{-1}$, discovery-level significance can be achieved for $e^+ e^- \to H^{++} H_1^- H_1^- \to 4\ell + \slashed{E}_T$, whereas $e^+ e^- \to H^{++} H_1^- W^- \to 4\ell + \slashed{E}_T$ typically requires higher luminosity to reach a comparable sensitivity. As expected, increasing $\delta$ induces a mild reduction in $\mathcal{Z}$, while the sensitivity degrades more noticeably at larger $\sqrt{s}$ due to the reduced signal cross-sections. We emphasize that systematic uncertainties are treated here in a simplified manner by assuming a flat fractional uncertainty on the total background~\cite{Cowan:2010js}.

				\begin{figure*}[!htb]
		\centering	
		\includegraphics[scale=0.35]{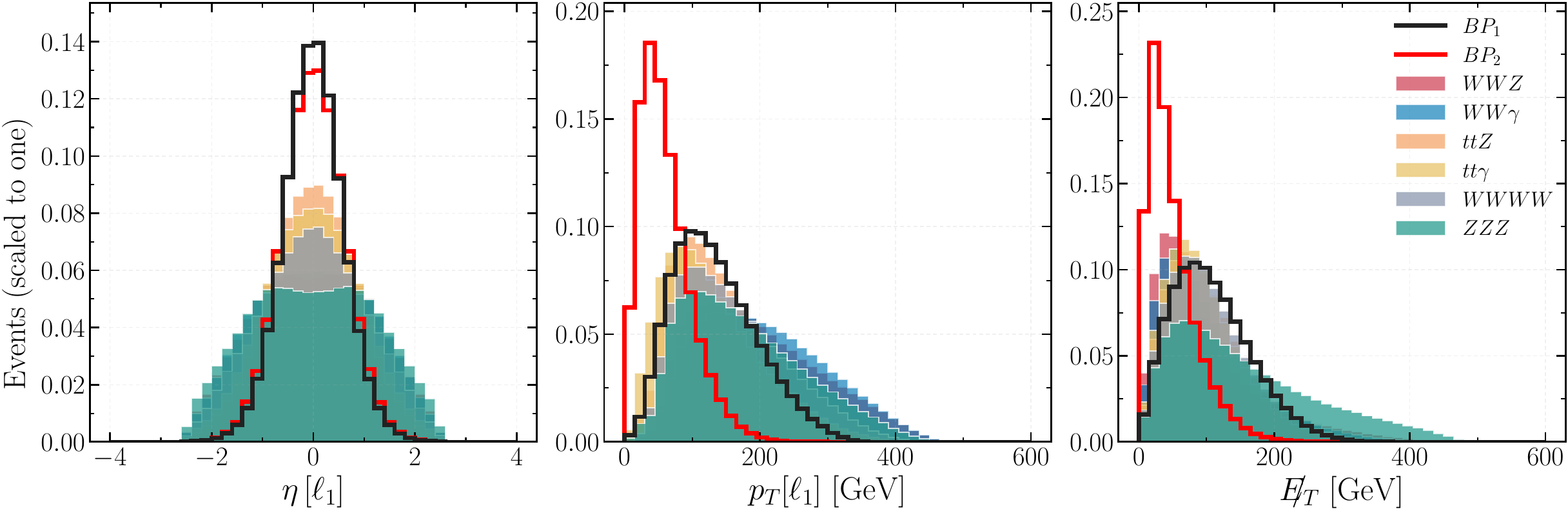}
	
\caption{Kinematic distributions for the $4\ell + \slashed{E}_T$ final state at $\sqrt{s}=1000$ GeV. The pseudorapidity $\eta(\ell_1)$ of the leading lepton (left), the transverse momentum $p_T(\ell_1)$ (middle), and the missing transverse energy $\slashed{E}_T$ (right).}
		\label{p1}
	\end{figure*}
\begin{figure*}[!htb]
	\centering	
	\includegraphics[scale=0.35]{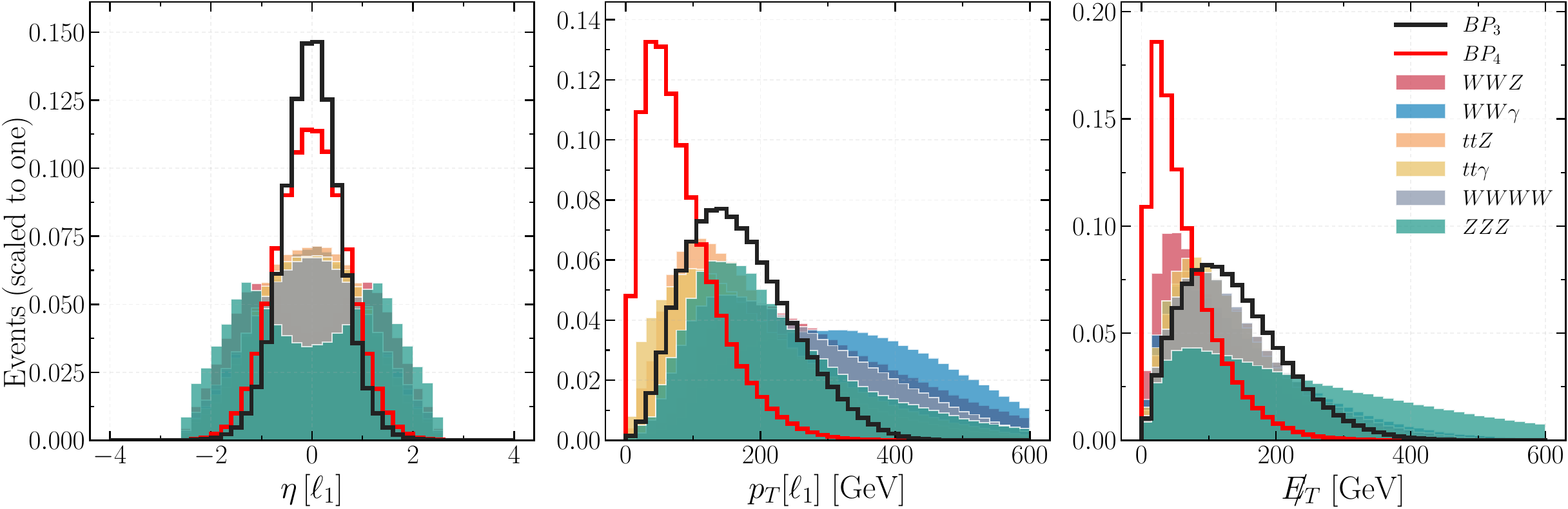}
\caption{Kinematic distributions for the $4\ell + \slashed{E}_T$ final state at $\sqrt{s}=1500$ GeV. The pseudorapidity $\eta(\ell_1)$ of the leading lepton (left), the transverse momentum $p_T(\ell_1)$ (middle), and the missing transverse energy $\slashed{E}_T$ (right).}
	\label{p2}
\end{figure*}
			\begin{table*}[!htb]
	\centering
	\renewcommand{\arraystretch}{1.3}
	\setlength{\tabcolsep}{45pt}
	\begin{adjustbox}{max width=\textwidth}
		\begin{tabular}{l c} 
			\hline \hline
			\textbf{Cuts}  & \textbf{Definition} \\  
			\hline \hline	
			\textbf{Trigger} & $N(b) \leq 1$ \\  	
			\hline
			\textbf{Cut-1} & $-0.9 < \eta [l_{1}] < 0.9$  \\   
			\hline
			\textbf{Cut-2} & $P_T[l_{1}] < 300$ GeV \\   
			\hline
			\textbf{Cut-3} & $\slashed{E_T} < 300$ GeV \\
			\hline \hline
		\end{tabular}
	\end{adjustbox}		
	\caption{Selection criteria applied in the signal-to-background analysis of the process 
		$e^+ e^- \to H^{\pm\pm} H_1^\mp W^{\mp} \to 4l +\slashed{E}_T$ 
		at $\sqrt{s} = 1000$ GeV for BP1.}
	\label{selection_cuts_HppHp1W_BP1}
\end{table*}
			\begin{table*}[!htb]
	\centering
	\renewcommand{\arraystretch}{1.3}
	\setlength{\tabcolsep}{45pt}
	\begin{adjustbox}{max width=\textwidth}
		\begin{tabular}{l c} 
			\hline \hline
			\textbf{Cuts}  & \textbf{Definition} \\  
			\hline \hline	
			\textbf{Trigger} & $N(b) \leq 1$ \\  	
			\hline
			\textbf{Cut-1} & $-1.6 < \eta [l_{1}] < 1.6$  \\   
			\hline
			\textbf{Cut-2} & $P_T[l_{1}] < 200$ GeV \\   
			\hline
			\textbf{Cut-3} & $\slashed{E_T} < 200$ GeV \\
			\hline \hline
		\end{tabular}
	\end{adjustbox}		
	\caption{Selection criteria applied in the signal-to-background analysis of the process 
		$e^+ e^- \to H^{\pm\pm} H_1^\mp H_1^{\mp} \to 4l +\slashed{E}_T$ 
		at $\sqrt{s} = 1000$ GeV for BP2.}
	\label{selection_cuts_HppHp1Hp1_BP2}
\end{table*}
			\begin{table*}[t]
	\centering
	\renewcommand{\arraystretch}{1.3}
	\setlength{\tabcolsep}{45pt}
	\begin{adjustbox}{max width=\textwidth}
		\begin{tabular}{l c} 
			\hline \hline
			\textbf{Cuts}  & \textbf{Definition} \\  
			\hline \hline	
			\textbf{Trigger} & $N(b) \leq 1$ \\  	
			\hline
			\textbf{Cut-1} & $-0.9 < \eta [l_{1}] < 0.9$  \\   
			\hline
			\textbf{Cut-2} & $P_T[l_{1}] < 400$ GeV \\   
			\hline
			\textbf{Cut-3} & $\slashed{E_T} < 400$ GeV \\
			\hline \hline
		\end{tabular}
	\end{adjustbox}		
	\caption{Selection criteria applied in the signal-to-background analysis of the process $e^+ e^- \to H^{\pm\pm} H_1^\mp W^{\mp} \to 4l +\slashed{E}_T$ at $\sqrt{s} = 1500$ GeV for BP3.}
	\label{selection_cuts_HppHp1W_BP3}
\end{table*}
\begin{table*}[t]
	\centering
	\renewcommand{\arraystretch}{1.3}
	\setlength{\tabcolsep}{45pt}
	\begin{adjustbox}{max width=\textwidth}
		\begin{tabular}{l c} 
			\hline \hline
			\textbf{Cuts}  & \textbf{Definition} \\  
			\hline \hline	
			\textbf{Trigger} & $N(b) \leq 1$ \\  	
			\hline
			\textbf{Cut-1} & $-1.4 < \eta [l_{1}] < 1.4$  \\   
			\hline
			\textbf{Cut-2} & $P_T[l_{1}] < 300$ GeV \\   
			\hline
			\textbf{Cut-3} & $\slashed{E_T} < 300$ GeV \\
			\hline \hline
		\end{tabular}
	\end{adjustbox}		
	\caption{Selection criteria applied in the signal-to-background analysis of the process 
		$e^+ e^- \to H^{\pm\pm} H_1^\mp H_1^{\mp} \to 4l +\slashed{E}_T$ 
		at $\sqrt{s} = 1500$ GeV for BP4.}
	\label{selection_cuts_HppHp1Hp1_BP4}
\end{table*}
\begin{table*}[ht!]
	\centering
\hspace*{-1.0truecm}
	\setlength{\tabcolsep}{8pt}
	\renewcommand{\arraystretch}{1.2}
	\begin{tabular}{l c c c c c c c}
		\hline\hline
		\multirow{2}{*}{\textbf{Cuts}} & \textbf{Signal} & \multicolumn{6}{c}{\textbf{Backgrounds}} \\ 
		\cline{2-2} \cline{3-8} 
		& BP1 & $W^+W^-Z$ & $W^+W^-\gamma$& $t\bar tZ$ & $t\bar t\gamma$ & $W^+W^-W^+W^-$ & $ZZZ$\\
		\hline\hline
		Basic cut  
		& $1.00\times10^{-1}$ 
		& $1.22\times10^{-1}$ 
		& $1.98\times10^{-2}$ 
		& $1.07\times10^{-2}$ 
		& $2.08\times10^{-3}$ 
		& $1.26\times10^{-3}$ 
		& $1.04\times10^{-3}$ \\
		
		Tagger   
		& $9.99\times10^{-2}$ 
		& $1.221\times10^{-1}$ 
		& $1.979\times10^{-2}$ 
		& $2.44\times10^{-3}$ 
		& $4.58\times10^{-4}$ 
		& $1.26\times10^{-3}$ 
		& $1.04\times10^{-3}$ \\
		
		Cut-1   
		& $8.75\times10^{-2}$ 
		& $6.34\times10^{-2}$ 
		& $1.03\times10^{-2}$ 
		& $1.71\times10^{-3}$ 
		& $3.04\times10^{-4}$ 
		& $7.52\times10^{-4}$ 
		& $4.96\times10^{-4}$ \\
		
		Cut-2   
		& $8.68\times10^{-2}$ 
		& $4.48\times10^{-2}$ 
		& $7.04\times10^{-3}$ 
		& $1.60\times10^{-3}$ 
		& $2.83\times10^{-4}$ 
		& $6.77\times10^{-4}$ 
		& $3.83\times10^{-4}$ \\
		
		Cut-3   
		& $8.66\times10^{-2}$ 
		& $4.27\times10^{-2}$ 
		& $6.88\times10^{-3}$ 
		& $1.59\times10^{-3}$ 
		& $2.81\times10^{-4}$ 
		& $6.73\times10^{-4}$ 
		& $3.14\times10^{-4}$ \\
		\hline
		\textbf{Total efficiencies (\%)} 
		& \textbf{86.6} 
		& \textbf{34.9} 
		& \textbf{34.7} 
		& \textbf{14.8} 
		& \textbf{13.5} 
		& \textbf{53.3} 
		& \textbf{30.3} \\
		\hline\hline
	\end{tabular}
	\caption{Cut flow of the cross-sections (in fb) for the signal and SM backgrounds at $\sqrt{s}=1000$~GeV using  BP1.}
	\label{cutft_HppHp1W_BP1}
\end{table*}

\begin{table*}[ht!]
	\centering
\hspace*{-1.0truecm}
	\setlength{\tabcolsep}{8pt}
	\renewcommand{\arraystretch}{1.2}
	\begin{tabular}{l c c c c c c c}
		\hline\hline
		\multirow{2}{*}{\textbf{Cuts}} & \textbf{Signal} & \multicolumn{6}{c}{\textbf{Backgrounds}} \\ 
		\cline{2-2} \cline{3-8} 
		& BP2 & $W^+W^-Z$ & $W^+W^-\gamma$& $t\bar tZ$ & $t\bar t\gamma$ & $W^+W^-W^+W^-$ & $ZZZ$\\
		\hline\hline
		Basic cut  
		& $7.94\times10^{-1}$ 
		& $1.22\times10^{-1}$ 
		& $1.98\times10^{-2}$ 
		& $1.07\times10^{-2}$ 
		& $2.08\times10^{-3}$ 
		& $1.26\times10^{-3}$ 
		& $1.04\times10^{-3}$ \\
		
		Tagger   
		& $7.943\times10^{-1}$ 
		& $1.221\times10^{-1}$ 
		& $1.979\times10^{-2}$ 
		& $2.44\times10^{-3}$ 
		& $4.58\times10^{-4}$ 
		& $1.26\times10^{-3}$ 
		& $1.04\times10^{-3}$ \\
		
		Cut-1   
		& $7.76\times10^{-1}$ 
		& $1.01\times10^{-1}$ 
		& $1.65\times10^{-2}$ 
		& $2.28\times10^{-3}$ 
		& $4.20\times10^{-4}$ 
		& $1.10\times10^{-3}$ 
		& $8.25\times10^{-4}$ \\
		
		Cut-2   
		& $7.74\times10^{-1}$ 
		& $4.86\times10^{-2}$ 
		& $7.08\times10^{-3}$ 
		& $1.69\times10^{-3}$ 
		& $3.15\times10^{-4}$ 
		& $7.42\times10^{-4}$ 
		& $4.55\times10^{-4}$ \\
		
		Cut-3   
		& $7.72\times10^{-1}$ 
		& $4.30\times10^{-2}$ 
		& $6.54\times10^{-3}$ 
		& $1.56\times10^{-3}$ 
		& $2.94\times10^{-4}$ 
		& $7.11\times10^{-4}$ 
		& $3.19\times10^{-4}$ \\
		\hline
		\textbf{Total efficiencies (\%)} 
		& \textbf{97.2} 
		& \textbf{35.1} 
		& \textbf{33.0} 
		& \textbf{14.5} 
		& \textbf{14.1} 
		& \textbf{56.3} 
		& \textbf{30.8} \\
		\hline\hline
	\end{tabular}
	\caption{Cut flow of the cross-sections (in fb) for the signal and SM backgrounds at $\sqrt{s}=1000$~GeV using  BP2.}
	\label{cutft_HppHp1Hp1_BP2}
\end{table*}

\begin{table*}[ht!]
	\centering
\hspace*{-1.0truecm}
	\setlength{\tabcolsep}{8pt}
	\renewcommand{\arraystretch}{1.2}
	\begin{tabular}{l c c c c c c c}
		\hline\hline
		\multirow{2}{*}{\textbf{Cuts}} & \textbf{Signal} & \multicolumn{6}{c}{\textbf{Backgrounds}} \\ 
		\cline{2-2} \cline{3-8} 
		& BP3 & $W^+W^-Z$ & $W^+W^-\gamma$& $t\bar tZ$ & $t\bar t\gamma$ & $W^+W^-W^+W^-$ & $ZZZ$\\
		\hline\hline
		Basic cut  
		& $5.87\times10^{-2}$ 
		& $8.30\times10^{-2}$ 
		& $1.86\times10^{-2}$ 
		& $1.46\times10^{-3}$ 
		& $3.03\times10^{-4}$ 
		& $1.66\times10^{-3}$ 
		& $8.09\times10^{-4}$ \\
		
		Tagger   
		& $5.873\times10^{-2}$ 
		& $8.30\times10^{-2}$ 
		& $1.858\times10^{-2}$ 
		& $1.46\times10^{-3}$ 
		& $3.03\times10^{-4}$ 
		& $1.66\times10^{-3}$ 
		& $8.09\times10^{-4}$ \\
		
		Cut-1   
		& $5.25\times10^{-2}$ 
		& $3.47\times10^{-2}$ 
		& $1.08\times10^{-2}$ 
		& $8.72\times10^{-4}$ 
		& $1.77\times10^{-4}$ 
		& $9.33\times10^{-4}$ 
		& $2.81\times10^{-4}$ \\
		
		Cut-2   
		& $5.23\times10^{-2}$ 
		& $1.89\times10^{-2}$ 
		& $5.12\times10^{-3}$ 
		& $7.44\times10^{-4}$ 
		& $1.39\times10^{-4}$ 
		& $6.99\times10^{-4}$ 
		& $1.96\times10^{-4}$ \\
		
		Cut-3   
		& $5.22\times10^{-2}$ 
		& $1.79\times10^{-2}$ 
		& $4.91\times10^{-3}$ 
		& $7.30\times10^{-4}$ 
		& $1.37\times10^{-4}$ 
		& $6.91\times10^{-4}$ 
		& $1.35\times10^{-4}$ \\
		\hline
		\textbf{Total efficiencies (\%)} 
		& \textbf{88.7} 
		& \textbf{21.7} 
		& \textbf{26.4} 
		& \textbf{10.2} 
		& \textbf{9.5} 
		& \textbf{41.5} 
		& \textbf{16.7} \\
		\hline\hline
	\end{tabular}
	\caption{Cut flow of the cross-sections (in fb) for the signal and SM backgrounds at $\sqrt{s}=1500$~GeV using BP3.}
	\label{cutft_HppHp1W_BP3}
\end{table*}

\begin{table*}[ht!]
	\centering
\hspace*{-1.0truecm}
	\setlength{\tabcolsep}{8pt}
	\renewcommand{\arraystretch}{1.2}
	\begin{tabular}{l c c c c c c c}
		\hline\hline
		\multirow{2}{*}{\textbf{Cuts}} & \textbf{Signal} & \multicolumn{6}{c}{\textbf{Backgrounds}} \\ 
		\cline{2-2} \cline{3-8} 
		& BP4 & $W^+W^-Z$ & $W^+W^-\gamma$& $t\bar tZ$ & $t\bar t\gamma$ & $W^+W^-W^+W^-$ & $ZZZ$\\
		\hline\hline
		Basic cut  
		& $1.947\times10^{-1}$ 
		& $8.30\times10^{-2}$ 
		& $1.858\times10^{-2}$ 
		& $1.46\times10^{-3}$ 
		& $3.03\times10^{-4}$ 
		& $1.66\times10^{-3}$ 
		& $8.09\times10^{-4}$ \\
		
		Tagger   
		& $1.947\times10^{-1}$ 
		& $8.30\times10^{-2}$ 
		& $1.858\times10^{-2}$ 
		& $1.46\times10^{-3}$ 
		& $3.03\times10^{-4}$ 
		& $1.66\times10^{-3}$ 
		& $8.09\times10^{-4}$ \\
		
		Cut-1   
		& $1.83\times10^{-1}$ 
		& $5.79\times10^{-2}$ 
		& $1.49\times10^{-2}$ 
		& $1.21\times10^{-3}$ 
		& $2.45\times10^{-4}$ 
		& $1.30\times10^{-3}$ 
		& $5.02\times10^{-4}$ \\
		
		Cut-2   
		& $1.83\times10^{-1}$ 
		& $2.44\times10^{-2}$ 
		& $4.45\times10^{-3}$ 
		& $9.20\times10^{-4}$ 
		& $1.65\times10^{-4}$ 
		& $7.66\times10^{-4}$ 
		& $3.09\times10^{-4}$ \\
		
		Cut-3   
		& $1.83\times10^{-1}$ 
		& $2.19\times10^{-2}$ 
		& $4.03\times10^{-3}$ 
		& $8.53\times10^{-4}$ 
		& $1.55\times10^{-4}$ 
		& $7.37\times10^{-4}$ 
		& $1.87\times10^{-4}$ \\
		\hline
		\textbf{Total efficiencies (\%)} 
		& \textbf{93.6} 
		& \textbf{26.4} 
		& \textbf{21.6} 
		& \textbf{12.0} 
		& \textbf{10.8} 
		& \textbf{44.3} 
		& \textbf{23.0} \\
		\hline\hline
	\end{tabular}
	\caption{Cut flow of the cross-sections (in fb) for the signal and SM backgrounds at $\sqrt{s}=1500$~GeV using BP4.}
	\label{cutft_HppHp1Hp1_BP4}
\end{table*}

\begin{table*}[ht!]
	\centering
	\setlength{\tabcolsep}{6pt}
	\renewcommand{\arraystretch}{1.25}
	\begin{tabular}{c c c c c c c}
		\hline\hline
		BP & Process & $\sqrt{s}$ [GeV] & $\delta$ [\%] &
		$\mathcal{L}=500~\mathrm{fb}^{-1}$ &
		$\mathcal{L}=1000~\mathrm{fb}^{-1}$ &
		$\mathcal{L}=1500~\mathrm{fb}^{-1}$ \\
		\hline\hline
		\multirow{2}{*}{BP1} 
		& \multirow{2}{*}{$e^+ e^- \to H^{\pm\pm} H_1^\mp W^{\mp}\to 4l +\slashed{E}_T$}
		& \multirow{2}{*}{1000} & 5  & 4.65 & \underbar{6.03} & \underbar{6.86} \\
		&  &  & 10 & 3.72 & 4.32 & 4.60 \\
		\hline
		\multirow{2}{*}{BP2} 
		& \multirow{2}{*}{$e^+ e^- \to H^{\pm\pm} H_1^\mp H_1^{\mp}\to 4l +\slashed{E}_T$}
		& \multirow{2}{*}{1000} & 5  & \underbar{25.62} & \underbar{32.15} & \underbar{35.84} \\
		&  &  & 10 & \underbar{19.14} & \underbar{21.59} & \underbar{22.68} \\
		\hline
		\multirow{2}{*}{BP3} 
		& \multirow{2}{*}{$e^+ e^- \to H^{\pm\pm} H_1^\mp W^{\mp}\to 4l +\slashed{E}_T$}
		& \multirow{2}{*}{1500} & 5  & 3.36 & 4.46 & \underbar{5.15} \\
		&  &  & 10 & 2.82 & 3.38 & 3.66  \\
		\hline
		\multirow{2}{*}{BP4} 
		& \multirow{2}{*}{$e^+ e^- \to H^{\pm\pm} H_1^\mp H_1^{\mp}\to 4l +\slashed{E}_T$}
		& \multirow{2}{*}{1500} & 5  & \underbar{9.71} & \underbar{12.71} & \underbar{14.56} \\
		&  &  & 10 & \underbar{7.93} & \underbar{9.34} & \underbar{10.02} \\
		\hline\hline
	\end{tabular}
	\caption{Significances $\mathcal{Z}$ for all BPs. Systematic uncertainties $\delta=5\%$ and $10\%$ are included at energies $\sqrt{s}=1000$ and $1500$~GeV for integrated luminosities of $\mathcal{L}=500$, $1000$, and $1500~\mathrm{fb}^{-1}$. The underlined results refer to the case of a clear discovery.}
	\label{Significance_AllBP}
\end{table*}

\section{CONCLUSION}
\vspace{6pt}
We have investigated the discovery prospects for a doubly charged Higgs boson in the type-X configuration  of the 2HDMcT framework at future $e^+e^-$ colliders, concentrating on the $2\to3$ body production channels $e^+e^- \to H^{\pm\pm} H_1^\mp H_1^{\mp}$ and $e^+e^- \to H^{\pm\pm} H_1^\mp W^{\mp}$. We have done so by carrying out a detailed detector level MC simulation of the ensuing $4\ell +\slashed{E}_T$ signal ($\ell=e,\mu$) and dominant SM backgrounds.
The BPs considered were required to satisfy theoretical requirements of self-consistency and to comply with current experimental limits.

We have found that these production modes can substantially exceed the yield of the conventional  ($2\to2$ body) pair-production process $e^+e^- \to H^{++}H^{--}$ followed by the decays $H^{\pm\pm}\to H_1^\pm H_1^\pm$ and $H^{\pm\pm}\to H_1^\pm W^\pm$, even when the latter are kinematically open, over sizable regions of the parameter space of our BSM scenario.

In short, our results establish $e^+e^- \to H^{\pm\pm} H_1^\mp H_1^{\mp}$ and $e^+e^- \to H^{\pm\pm} H_1^\mp W^{\mp}$ as powerful and, for some parameter configurations, leading probes of $H^{\pm\pm}$ states at lepton colliders, as our sophisticated signal-to-background study has demonstrated that a $5\sigma$ discovery is achievable through a very clean BSM signature 
for representative BPs at promoted CoM energies and integrated luminosities of a future $e^+e^-$ collider operating at the TeV scale. Furthermore, beam polarization is expected to enhance the signal cross sections and thereby improve the overall discovery potential.

We therefore encourage experimental analyses to depart from mainstream approaches leveraging $H^{\pm\pm}$ 
pair-production and decay to test triplet Higgs models in $e^+e^-$ settings. In fact, we finally emphasize that our choice of BSM scenario is by no means restrictive, as it was used for quantitatively illustrative purposes of the advocated new discovery channels.
\vspace{6pt}
\label{conlusion}	
\section*{ACKNOWLEDGMENTS}
A. Arhrib is supported by the Arab Fund for economic and social development. M. Boukidi acknowledges the support of the Narodowe Centrum Nauki under OPUS Grant No. 2023/49/B/ST2/03862 as well as the use of the PALMA II high-performance computing cluster at the University of M\"unster, subsidized by the DFG (INST 211/667-1). K. Goure would like to thank CNRST/HPC-MARWAN for technical support. S. Moretti is supported in part through the NExT Institute and  STFC CG ST/X000583/1.

\end{document}